\documentclass[preprint,nofootinbib,amsmath,amssymb,aps,prd,superscriptaddress,a4paper]{revtex4-1}
\pdfoutput=1

\usepackage{graphicx}% Include figure files
\usepackage{dcolumn}% Align table columns on decimal point
\usepackage{bm}% bold math
\usepackage{slashed}
\usepackage{physics}
\usepackage{comment}
\usepackage{cancel}
\usepackage{subcaption}
\usepackage{caption} 
\usepackage[normalem]{ulem}
\usepackage[colorlinks=true,linkcolor=blue,citecolor=red,urlcolor=magenta]{hyperref}
\usepackage{ragged2e}

\usepackage[capitalize,nameinlink]{cleveref}

\begin{document}

\title{Large lepton asymmetry from axion inflation \\ and helium abundance hinted by ACT}

\author{Di Wu}
\affiliation{School of Fundamental Physics and Mathematical Sciences, Hangzhou Institute for Advanced Study, University of Chinese Academy of Sciences (HIAS-UCAS), 310024 Hangzhou, China}
\affiliation{Institute of Theoretical Physics, Chinese Academy of Sciences, Beijing 100190, China}
\affiliation{University of Chinese Academy of Sciences, Beijing 100049, China}

\author{Yifan Hu}
\affiliation{School of Fundamental Physics and Mathematical Sciences, Hangzhou Institute for Advanced Study, University of Chinese Academy of Sciences (HIAS-UCAS), 310024 Hangzhou, China}
\affiliation{Institute of Theoretical Physics, Chinese Academy of Sciences, Beijing 100190, China}
\affiliation{University of Chinese Academy of Sciences, Beijing 100049, China}

\author{Kohei Kamada}
\affiliation{School of Fundamental Physics and Mathematical Sciences, Hangzhou Institute for Advanced Study, University of Chinese Academy of Sciences (HIAS-UCAS), 310024 Hangzhou, China}
\affiliation{International Centre for Theoretical Physics Asia-Pacific (ICTP-AP), Hangzhou/Beijing, China}
\affiliation{Research Center for the Early Universe, The University of Tokyo, Bunkyo-ku, Tokyo 113-0033, Japan}

\begin{abstract}
    The generation of helical magnetic fields and the associated chiral asymmetry via the chiral anomaly is a generic feature in pseudoscalar inflation. 
    In the presence of a Chern--Simons coupling between the inflaton and a U(1) gauge field, the homogeneous evolution of the inflaton induces a tachyonic instability in one circular polarization of the gauge field, resulting in the production of helical magnetic fields. 
    In this work, we show that, in the case of a gauged lepton flavor symmetry, U(1)$_{L_i-L_j}$, this mechanism can lead to the generation of a sizable lepton asymmetry.
    In a simple setup, however, the resulting lepton asymmetry  is typically too small  to have an observational consequences, even setting aside constraints from baryon overproduction via sphaleron processes, due to the backreaction of the produced gauge fields and fermions on the inflationary dynamics. 
    We demonstrate that this limitation can be overcome by implementing a mechanism to suppress fermion production during inflation.  
    As a result, a much larger lepton asymmetry can be generated from the subsequent decay of magnetic helicity.  
    Remarkably, for the gauged U(1)$_{L_\mu-L_\tau}$ symmetry, the generated asymmetry  can be sufficiently large to suppress the primordial helium abundance, as may be inferred from recent cosmic microwave background observations by ACT. 
\end{abstract}

\maketitle

\section{Introduction}

The $\Lambda$ cold dark matter ($\Lambda$CDM) model is the standard paradigm for describing the present Universe. 
A wide range of cosmological observations, including cosmic microwave background (CMB) and the the large-scale structure of the Universe, are well explained within this framework~\cite{Peebles:2024txt,Planck:2018vyg}. 
In particular, these observations strongly favor a nearly scale-invariant spectrum of primordial density perturbations, which is a generic prediction of inflation~\cite{Mukhanov:1981xt,Hawking:1982cz,Starobinsky:1982ee,Guth:1982ec,Bardeen:1983qw,Lyth:1998xn}. 
While CMB measurements up to Planck~\cite{Planck:2018jri} are consistent with the spectral index predicted by some of the simplest and well-motivated inflationary models, such as the $R^2$-model~\cite{Starobinsky:1980te} and Higgs inflation~\cite{Spokoiny:1984bd,Futamase:1987ua,Cervantes-Cota:1995ehs,Bezrukov:2007ep},  recent results from the Atacama Cosmology Telescope (ACT) indicate a preference for a relatively larger scalar spectral index~\cite{ACT:2025fju}, 
for which these representative models become mildly disfavored. 
See also results from the South Pole Telescope (SPT)~\cite{SPT-3G:2025bzu}. 

Of course, there is no fundamental reason to restrict ourselves to these simplest inflationary models, 
and one may instead consider models that predict a slightly larger spectral index. 
However, it is worth noting that some regions of parameter space that were disfavored by Planck become favored by ACT, 
which may point toward cosmological scenarios beyond the minimal $\Lambda$CDM framework~\cite{ACT:2025tim}. 
Indeed, ACT analyses performed within extended cosmological models indicate a preference for a lower helium abundance, $Y_p$, and a smaller effective number of relativistic species, $N_\mathrm{eff}$, while remaining compatible with the predictions of the simplest inflationary models. 
Since both $Y_p$ and $N_\mathrm{eff}$ affect the Silk damping scale of primordial perturbations~\cite{Silk:1967kq,Dodelson:2020bqr}, these results suggest that the ACT data may be interpreted as indicating a weakening of the exponential damping at small scales, 
rather than a genuine enhancement of the power-law tilt. 
Although the shift in the spectral index becomes milder when the recent results of SPT are included, they still point toward a preference for a smaller helium abundance~\cite{SPT-3G:2025bzu}. 
Therefore, it is well motivated to explore cosmological scenarios that lead to a reduced helium abundance, as well as a modified effective number of neutrino species. 
Note that the primordial helium abundance favored by these CMB measurements is somewhat lower than the value quoted  in the Particle Data Group (PDG) 2024~\cite{ParticleDataGroup:2024cfk}. Nevertheless, recent measurements by the EMPRESS survey indicate values that are compatible with the former~\cite{Yanagisawa:2025mgx}. 

One well-motivated  way to reduce the primordial helium abundance is to generate a sizable electron neutrino asymmetry. 
Owing to the chemical equilibrium of the weak interaction around the epoch of Big Bang Nucleosynthesis (BBN), a positive electron neutrino asymmetry reduces the neutron-to-proton ratio, thereby leading to a lower helium abundance. 
In general, however, a large electron neutrino asymmetry implies the existence of a large primordial lepton asymmetry. 
If such a lepton asymmetry is generated before the electroweak phase transition (EWPT), it is partially converted into baryon asymmetry~\cite{tHooft:1976rip,tHooft:1976snw} through the sphaleron process~~\cite{Manton:1983nd,Klinkhamer:1984di,Kuzmin:1985mm,Arnold:1987mh}, resulting in an overproduction of baryon asymmetry of the Universe.  
Therefore, viable scenarios require a mechanism that generates a sizable lepton asymmetry after the EWPT but before BBN. See Refs.~\cite{Casas:1997gx,McDonald:1999in,Kawasaki:2002hq,Yamaguchi:2002vw,Takahashi:2003db,Asaka:2005pn,Harigaya:2019uhf,Gelmini:2020ekg,Kawasaki:2022hvx,Borah:2022uos,Bhandari:2023wit,ChoeJo:2023cnx,Borah:2024xoa,Bae:2025gpc} for such attempts~\footnote{Another possibility is to generate lepton flavor asymmetry~\cite{March-Russell:1999hpw,Shu:2006mm, Gu:2010dg, Mukaida:2021sgv,Akita:2025zvq}. Since the neutrino oscillation does not erase the lepton flavor asymmetry completely until BBN~\cite{Dolgov:2002ab,Pastor:2008ti,Froustey:2021azz,Froustey:2024mgf,Domcke:2025lzg}, a large electron neutrino asymmetry can remain if we have sufficiently large primordial lepton flavor asymmetry.}.  

Mechanisms that generate a large particle asymmetry are, in general, severely constrained by experimental bounds on CP violation if the CP violation is encoded in particle properties. 
Consequently, scenarios in which CP violation is provided by cosmological dynamics are particularly promising. 
Baryogenesis from axion inflation~\cite{Anber:2015yca,Adshead:2016iae,Jimenez:2017cdr,Domcke:2018eki,Domcke:2019mnd,Domcke:2022kfs} is one of the representative mechanisms in this class. 
If pseudoscalar inflaton has a Chern--Simons coupling to the U(1) hypercharge gauge field of the Standard Model of particle physics (SM), helical hypermagnetic fields are generated during inflation~\cite{Turner:1987bw,Garretson:1992vt,Anber:2006xt}. 
Through the chiral anomaly inherent in the SM~\cite{tHooft:1976rip,tHooft:1976snw}, a baryon asymmetry is produced simultaneously~\cite{Domcke:2018eki,Domcke:2019mnd}, together with additional baryon number generation driven by subsequent  cosmological dynamics~\cite{Giovannini:1997gp,Giovannini:1997eg,Fujita:2016igl,Kamada:2016eeb,Kamada:2016cnb,Domcke:2022kfs}. 
While regions of parameter space leading to the successful baryogenesis have been identified (see, however, recent discussions in Refs.~\cite{Hamada:2025cwu,Fukuda:2025nmc}), a larger asymmetry can in principle be obtained by assuming a stronger Chern--Simons coupling. 

In this study, we investigate the generation of  a large lepton asymmetry from axion inflation. 
In the presence of the Chern--Simons coupling between the inflaton and U(1) hypercharge gauge field, it is difficult to obtain a sufficiently large lepton asymmetry without simultaneously overproducing baryon asymmetry. 
Motivated by this, we consider a gauged lepton flavor symmetry, U(1)$_{L_i-L_j}$, where $L_i$ and $L_j$ denote the lepton numbers of the $i$-th and $j$-th generations, respectively, following  a similar spirit of Ref.~\cite{Fukuda:2024pkh}. 
Once gauged, the lepton number is anomalously broken. 
As a consequence, the generation of helical U(1)$_{L_i-L_j}$ magnetic fields during inflation is accompanied by the production of the lepton number asymmetry in the $i$ and $j$-th generation. 
We show that, even setting aside the constraint from baryon overproduction, there exists an upper bound on the achievable asymmetry arising from the backreaction of the produced gauge fields and fermions on the inflationary dynamics. 
As a result, the lepton asymmetry generated during inflation cannot be large enough to reduce the primordial helium abundance preferred by ACT and SPT. 
In contrast, if we implement a mechanism that effectively switches off the fermion (and hence lepton number) production during inflation, 
a sufficiently large lepton asymmetry can instead be generated  at the time of symmetry breaking after the EWPT and before BBN.  We find that this can be realized in the case of gauged U(1)$_{L_\mu-L_\tau}$ symmetry. 
Remarkably, the viable parameter region is  also compatible with explanations of   the possible muon $g-2$ anomaly~\cite{Aoyama:2020ynm,Muong-2:2021ojo,Muong-2:2024hpx}. 

This paper is organized as follows. 
In Sec.~\ref{sec:helium}, we review the current status of CMB observations and discuss how a large lepton asymmetry can explain these results. 
In Sec.~\ref{sec:MGwithSE}, we study magnetogenesis and lepton number generation during axion inflation with a gauged lepton flavor U(1) symmetry. 
We identify an upper bound of the resulting asymmetry arising from the backreaction of the produced gauge fields and fermions on the inflationary dynamics. 
In Sec.~\ref{sec:MGwithoutSE}, we construct a model in which  fermion production during inflation is effectively swithched off, and investigate the subsequent cosmic history to examine whether a sizable lepton asymmetry can be successfully generated. 
We find that a viable parameter region that can accommodate the ACT results 
in the case of a gauged U(1)$_{L_\mu-L_\tau}$ symmetry. 
Section~\ref{sec:summary} is devoted to the conclusion and discussion. 

\section{Cosmologically suggested Helium abundance and lepton flavor asymmetry \label{sec:helium}}

We begin by reviewing the interpretation of the recent results from ACT~\cite{ACT:2025fju} 
and explining why a large lepton asymmetry can naturally account for them. 
The $\Lambda$CDM analysis of the ACT data, combined with Planck data, CMB lensing from ACT and Planck, and baryon acoustic oscillation measurements from DESI DR1 (P-ACT-LB), indicates a preference for a slightly larger scalar spectral index, $n_s = 0.974 \pm 0.003$, compared to the value reported by Planck alone, $n_s = 0.965 \pm 0.004$~\cite{Planck:2018jri}.
As a consequence, inflationary models such as the $R^2$-model~\cite{Starobinsky:1980te} and Higgs inflation~\cite{Spokoiny:1984bd,Futamase:1987ua,Cervantes-Cota:1995ehs,Bezrukov:2007ep}, 
which were strongly favored by Planck, become mildly disfavored. 
One possible approach is simply to consider inflation models that yields a slightly larger spectral index. 

However, 
ACT analyses with extended cosmological models reveal a preference for a lower helium abundance $Y_p$ and a smaller effective number of neutrino species $N_\mathrm{eff}$~\cite{ACT:2025tim}
while still favoring a lower spectral index.
This feature can be understood as follows. 
The ACT data exhibit a mild enhancement of the CMB anisotropy power spectra at high multipoles relative to the predictions of the $\Lambda$CDM model with parameters favored by the Planck data. 
Within the $\Lambda$CDM model, the only way to enhance power at high multipoles is to increase the scalar spectral index, 
which modifies the spectrum in a power-law manner. 
In contrast, the primordial helium abundance and the presence of additional relativistic species, parameterized by $N_\mathrm{eff}$, control the Silk damping scale~\cite{Silk:1967kq,Dodelson:2020bqr}, thereby altering the CMB angular power spectra exponentially. 
From this perspective, the ACT data may be more naturally interpreted as indicating a weaker exponential damping, rather than a weaker power-law decay, in the high-$\ell$ region of the spectra.  
In this study, we pursue this possibility, namely that the ACT data favor a shorter Silk damping scale associated with a reduced helium abundance. 

Let us briefly review the physical mechanism by which a lower helium abundance enhances the CMB angular power spectra at high-$\ell$.
Silk damping refers to the suppression of the photon energy fluctuation around the recombination epoch due to the Thomson scattering between the photon and baryons in the primordial fluid. 
As a consequence, the CMB temperature power spectra decay exponentially at high wavenumber with a characteristic scale~\cite{Silk:1967kq,Dodelson:2020bqr}, 
\begin{equation}
    q_\mathrm{Silk}^{-2}\equiv  \int_0^t \mathrm{d}t^\prime \frac{1}{6 a^2 \sigma_T {\bar n_e}} \left[\frac{16}{15 (1+R)}+\frac{R^2}{(1+R)^2} \right]  \simeq \frac{8}{45} \int_0^t \frac{\mathrm{d}t^\prime}{a^2  \sigma_T {\bar n}_e}, 
\end{equation}
where $a$ is the scale factor, $n_e$ is the electron number density with the bar denoting the average, and $\sigma_T$ is the Thomson scattering cross section. 
$R=3{\bar \rho_B}/4 {\bar \rho_\gamma}$ represents the fraction between the baryon and photon energy densities, which is small around the recombination. 
Since the integrand receives its dominant contribution from the recombination epoch, the value of ${\bar n_e}$ around that time determines the Silk damping scale. This leads to the approximate relation
\begin{equation}
    q_\mathrm{Silk}^{-2} \propto \left.(H \sigma_T {\bar n_e})^{-1} \right|_\mathrm{recombination}. 
\end{equation}
Therefore, a shorter Silk damping scale, $q_\mathrm{Silk}^{-1}$, which would explain the ACT data, 
can be achieved by a larger electron number density at recombination. 

Indeed, a larger electron number density at recombination can be realized by a lower primordial helium abundance. 
The physical picture is simple. A larger helium abundance means a larger neutron abundance during BBN. 
A larger number of neutrons means that more electrons have been bounded in neutrons. 
Conversely, a lower helium abundance leaves more baryons in the form of ionized hydrogen, thereby increasing the electron number density. 
Quantitatively, the mean free electron number density is given by
\begin{equation}
    {\bar n_e} = X \left(1-Y_p\right) \Omega_\mathrm{b}, 
\end{equation}
where $X$ denotes the ionization fraction, $Y_p \equiv 4 n_\mathrm{He}/n_\mathrm{b}$ is the helium mass fraction to the nucleon mass, and $\Omega_\mathrm{b}$ is the baryon density parameter. 
This relation directly illustrates how a smaller helium abundance leads to a larger electron density at recombination. 
\footnote{One may also expect a larger Hubble parameter could reduce the Silk damping scale and thus enhance power at high multipoles. 
However, a larger Hubble parameter simultaneously leads to the shift of the peaks of the CMB power spectra due to a smaller sound horizon. As a result, when the full shape of the spectra is taken into account, a smaller Hubble parameter is favored by the ACT data, which is the reason why smaller $n_\mathrm{eff}$ is favored.}

A natural way to realize a reduced helium abundance is to invoke a large lepton asymmetry, or more specifically, a sizable electron neutrino asymmetry.  
In standard BBN, the helium abundance is primarily determined by the neutron-to-proton ratio at the time when the weak interactions decouple, $T \simeq 0.8~\mathrm{MeV}$.
In the absence of significant lepton asymmetries, chemical equilibrium 
condition fixes the neutron-to-proton number density ratio as
\begin{equation}
    \left.\frac{n_n}{n_p}\right|_{T \simeq 0.8 \mathrm{MeV}} = \left.\exp \left[-\frac{\Delta m}{T} \right]\right|_{T \simeq 0.8 \mathrm{MeV}} \simeq \frac{1}{6}, \quad \Delta m \equiv m_n - m_p=1.293 \mathrm{MeV},  \label{nnnp1}
\end{equation}
where $m_n$ and $m_p$ denote the neutron and proton masses, respectively. 
After the decoupling of weak interactions, neutrons gradually decay,  
so that the neutron-to-proton ratio decreases to
\begin{equation}
     \left.\frac{n_n}{n_p}\right|_{T \simeq 0.1 \mathrm{MeV}} \simeq \frac{1}{7},   
\end{equation} 
by the time of helium formation, $T \simeq 0.1$ MeV. 
Most of the neutrons surviving until this epoch are incorporated into helium nuclei, 
which leads to an approximate primordial helium mass fraction
\begin{equation}
    Y_p \simeq 0.25. 
\end{equation}
In the presence of a large electron neutrino asymmetry, 
the chemical equilibrium condition between the neutron and protons is modified to
\begin{equation}
    \frac{n_n}{n_p} = \exp\left[-\frac{\Delta m+\mu_{\nu_e}}{T} \right], 
\end{equation}
where $\mu_{\nu_e}$ is the chemical potential of the electron neutrino. 
Note that, although neutrinos may carry a sizable asymmetry, the chemical potentials of protons and neutrons should be small, $\mu \sim 10^{-10} T$, reflecting the small baryon asymmetry of the Universe. 
The resulting shift of the neutron number directly translates into a change in the helium mass fraction. 
Analytically, we obtain a rough estimate, 
\begin{equation}
    Y_p \simeq 0.25 \times \left(1-\left.\xi_e\right|_{T \simeq 0.8 \mathrm{MeV}} \right), \quad \xi_e \equiv \frac{\mu_{\nu_e}}{T}. 
\end{equation}

For a more accurate estimate, we need to perform the numerical calculation of the BBN, taking into account all the relevant interactions during the process neglected in the above rough estimate. 
In the calculation with a public code \texttt{PRIMAT}~\cite{Pitrou:2018cgg}, it is estimated as
\begin{equation}
    Y_p = Y_p|_{\mu_{\nu_e}=0} \times \left(1-0.96\xi_e\right),  \quad Y_p|_{\mu_{\nu_e}=0} = 0.24709 \pm 0.00018,\label{YpPRIMAT}
\end{equation}
where the neutron life time is taken to be $\tau_n = 879.5(8)$s, and  $\mu_{\nu_e}/T$ is assumed to be constant. 
A recent study in Ref.~\cite{Domcke:2025jiy} also gives a fitting formula that includes the change of $N_\mathrm{eff}$ and 
the deviation of the neutrino temperature from the SM predictions $\delta_T$.  
This fitting formula and the public code associated with that work are used in our estimate of the helium abundance.

Let us now turn to a comparison to the observational results. 
The P-ACT-LB analysis within the $\Lambda$CDM+$N_\mathrm{eff}$+$Y_p$ model yields~\cite{ACT:2025tim} 
\begin{equation}
    Y_p= 0.227 \pm 0.014, \quad N_\mathrm{eff} = 3.14 \pm 0.25, \label{heliumNeffACT}
\end{equation}
which can be understood by the underlying physics discussed above. 
This value of helium abundance exhibits a discrepancy with the averaged helium abundance reported by the Particle Data Group (PDG) 2024, $Y_p= 0.245 \pm 0.003$~\cite{ParticleDataGroup:2024cfk}. 
However, the PDG value does not include recent measurements from the EMPRESS survey, which indicate a lower helium abundance, $Y_p= 0.2387^{+0.0036}_{-0.0031}$~\cite{Yanagisawa:2025mgx}, consistent with the present ACT result. 
The SPT collaboration has also released new data. The CMB-SPA analysis (a combination of SPT-3G D1, ACT DR6, and Planck) within the $\Lambda$CDM+$N_\mathrm{eff}$+$Y_p$ model gives~\cite{SPT-3G:2025bzu} 
\begin{equation}
    Y_p= 0.231 \pm 0.014, \quad N_\mathrm{eff} = 2.99^{+0.22}_{-0.26}. \label{heliumNeffSPT}
\end{equation}
Motivated by these results, in the following we investigate scenarios that lead to a helium abundance consistent with Eq.~\eqref{heliumNeffACT} as well as Eq.~\eqref{heliumNeffSPT}. 

By using the method (numerical calculation with the fitting formula) presented in Ref.~\cite{Domcke:2025jiy}, we find that 
the ACT results (Eq.~\eqref{heliumNeffACT}) can be explained if the  electron neutrino asymmetry at the time of BBN lines in the range 
\begin{equation}
    0.0248 \lesssim \xi_e \lesssim 0.151 , \label{eq:xieACT}
\end{equation}
 which corresponds to the first generation lepton-to-entropy ratio
\begin{equation}
    9.1\times10^{-4} \lesssim \frac{n_{L_e}}{s} \lesssim   5.3\times10^{-3},  \label{eq:nueasymACT}
\end{equation}
where $n_{L_e}$ and $s$ denote the first generation lepton number density and entropy density,respectively. 
Similarly, the SPT results (Eq.~\eqref{heliumNeffSPT}) can be accommodated if
\begin{equation}
    0.00812 \lesssim \xi_e \lesssim 0.132 , \label{eq:xieSPT}
\end{equation}
which corresponds to 
\begin{equation}
    3.0 \times 10^{-4} \lesssim \frac{n_{L_e}}{s} \lesssim   4.6\times10^{-3}.
    \label{eq:nueasymSPT}
\end{equation}
Although the precise numerical values may depend on the details of the BBN calculation,
in the following we adopt these ranges as benchmark targets to be achieved.

How can such a large electron neutrino asymmetry 
be generated in the early universe?
In the SM, the baryon-minus-lepton number $B-L$ is conserved, whereas $B+L$ asymmetry is anomalously broken~\cite{tHooft:1976rip,tHooft:1976snw}. In particular, the violation of $B+L$ efficiently proceeds
through the sphaleron process~\cite{Manton:1983nd,Klinkhamer:1984di,Arnold:1987mh} before the EWPT~\cite{Kuzmin:1985mm}. 
If baryogenesis takes place before the EWPT, the resulting total lepton asymmetry, and hence 
the electron neutrino asymmetry, 
is generically of the same order of magnitude as the baryon asymmetry, leading to
$n_L/s\sim10^{-10}$. 
Such a small  lepton asymmetry is therefore insufficient to account for the reduction of 
 the helium abundance discussed above.

One way to overcome this problem is to consider leptogenesis occurring after the EWPT (see, {\it e.g.,} 
\cite{Casas:1997gx,McDonald:1999in,Kawasaki:2002hq,Yamaguchi:2002vw,Takahashi:2003db,Asaka:2005pn,Harigaya:2019uhf,Gelmini:2020ekg,Kawasaki:2022hvx,Borah:2022uos,Bhandari:2023wit,ChoeJo:2023cnx,Borah:2024xoa,Bae:2025gpc}). In this low-temperature regime, the sphaleron rate is exponentially suppressed, and the sphaleron process effectively decouples~\cite{tHooft:1976rip,tHooft:1976snw}. The absence of this B+L-violating process 
shuts off the conversion channel between the lepton and baryon sectors. Consequently, leptogneesis scenarios 
operating after EWPT are exempt from the stringent constraints, $n_L/s \sim 10^{-10} (\ll 10^{-3})$, which applies to pre-EWPT scenarios. Such models can therefore, in principle, generate a sizable total lepton asymmetry without inducing a corresponding baryon asymmetry, for instance,
\begin{equation}
    \frac{n_{L_e}}{s} \sim \frac{n_{L_{\mu}}}{s} \sim\frac{n_{L_{\tau}}}{s} \sim \frac{n_L}{s}\sim 10^{-3}.
\end{equation}
In this study, we explore this direction by taking advantage of helical magnetogenesis from axion inflation. 

Before investigating a concrete model of leptogenesis, let us make a few remarks. 
When neutrino oscillations are ineffective, {\it i.e.}, at temperatures $T \gtrsim 10$ MeV, 
the lepton number is conserved separately for each flavor. 
One might then worry that a leptogenesis scenario which initially generates no asymmetry in the first generation cannot account for the reduction of helium abundance. 
However, neutrino oscillations become effective well before the onset of the BBN, thereby tending to equilibrate the lepton asymmetries among three flavors, $n_{L_e} \sim n_{L_\mu} \sim n_{L_\tau}$. 
This equilibration is known to be incomplete. 
In the case of an initial condition $\xi_\mu=\xi_\tau \equiv \xi^\mathrm{ini}$, which is the situation of interest in this study, 
we find, using the numerical code in Ref.~\cite{Domcke:2025jiy}, 
that
\begin{align}
    &\xi_e \simeq 0.7280 \xi^\mathrm{ini},~\text{for}~\text{Inverted Hierarchy} \nonumber \\
    &\xi_e \simeq 0.6576 \xi^\mathrm{ini},~\text{for}~\text{Normal Hierarchy} \label{eq:neoscillation}
\end{align}
at the time of BBN. 
In deriving Eqs.~\eqref{eq:xieACT}-- \eqref{eq:nueasymSPT}, we assumed this specific initial condition.
For different initial configurations, the resulting constraints are modified slightly. 
For definiteness, however, we shall take Eqs~\eqref{eq:nueasymACT} and \eqref{eq:nueasymSPT} as the target lepton asymmetries to be generated, allowing for an $\mathcal{O}(10)$ \% uncertainty.

\section{Axion inflation with gauged lepton flavor U(1) symmetry \label{sec:MGwithSE}}
Now we study the large lepton asymmetry generation from axion inflation. 
We first discuss the model based on a gauged lepton flavor symmetry and its constraints.
Specifically, we consider the gauged $U(1)_{L_i-L_j}$ symmetry, where the indices $i$ and $j$ represent the lepton flavors, $e, \mu$ and $\tau$.
In the limit of vanishing neutrino masses, the lepton flavor symmetry is exact and can thus be gauged.
The model is parameterized by the gauge coupling $g_{L_i-L_j}$ and the mass of the corresponding gauge boson $m_{Z'}$.
For gauge boson masses around MeV -- GeV scales, the main experimental constraints arise from beam-dump and fixed target experiments, as well as  collider searches~\cite{Bauer:2018onh}. 
As we will show below, 
in a simple setup there is an upper bound on the lepton asymmetry due to the backreaction arising from the produced gauge fields and fermions on the inflationary dynamics. 
In the next section, we will show that 
in the case of gauged U(1)$_{L_\mu-L_\tau}$ symmetry, 
there remains a viable parameter region for our purpose with a more complicated setup where fermions become heavy during inflation so that they are not produced during inflation and symmetry-breaking
scale lies below the electroweak scale.  
Interestingly, this parameter region can potentially account for the $(g-2)_\mu$ anomaly~\cite{Aoyama:2020ynm} (see, however, Ref.~\cite{Aliberti:2025beg}). 

Within the particle content of the SM, 
the system exhibits the chiral anomaly, 
\begin{equation}
    \partial_\mu J^\mu_{L_i} = \partial_\mu J^\mu_{L_j} = \frac{g_{{L_i}-{L_j}}^2}{16 \pi^2} X_{\mu\nu} {\tilde X}^{\mu \nu}, 
\end{equation}
where 
$X_{\mu\nu}$ is the field strength tensor of the U(1) $_{L_i-L_j}$ gauge field, $X_\mu$, and ${\tilde X}^{\mu\nu}$ is its dual. 
$J^\mu_{L_i}$ denotes the $i$-th lepton flavor current. 
While the lepton-flavor difference is anomaly free, 
\begin{equation}
    \partial_\mu J^\mu_{L_i-L_j}=0, 
\end{equation}
and is therefore gaugeable, the total lepton number is violated, 
\begin{equation}
    \partial_\mu J^\mu_{L}=\partial_\mu (J^\mu_{L_i} +J^\mu_{L_j} )=\frac{g_{{L_i}-{L_j}}^2}{8 \pi^2} X_{\mu\nu} {\tilde X}^{\mu \nu}.  
\end{equation}

Integrating the anomaly equation over the spatial volume yields 
\begin{equation}
    \frac{\mathrm{d}}{\mathrm{dt}} \left(Q_{L_i}-\frac{g_{{L_i}-{L_j}}^2}{8 \pi^2}  \mathcal{H}_X \right)=\frac{\mathrm{d}}{\mathrm{dt}} \left(Q_{L_j}-\frac{g_{{L_i}-{L_j}}^2}{8 \pi^2} \mathcal{H}_X \right)=0, \label{eq:csconservation}
\end{equation}
where $Q_{L_i}$ is the $i$-th lepton flavor number in the system, and 
\begin{equation}
 \mathcal{H}_X \equiv \int \mathrm{d}^3 x {\bm X} \cdot {\bm B}_X,
\end{equation}
is the magnetic helicity for the U(1)$_{L_i-L_j}$ gauge field. 
Here ${\bm X}$ is the spatial component of the gauge field
and ${\bm B}_X$ is its magnetic field. 
In the absence of the magnetic monopoles and boundary magnetic flux, 
the magnetic helicity is gauge invariant (see Ref.~\cite{Fukuda:2025nmc} for recent discussion). 

Equation~\eqref{eq:csconservation} therefore encodes two conservation laws of the summation of the $L_i-L_j$ magnetic helicity and the $i$- and $j$-th lepton-flavor charges
in the fermion sector, respectively.
Consequently, both lepton flavor charges, 
and hence an excess of $i$- and $j$-flavor neutrinos over their anti-particle (as well as the net lepton asymmetry) are expected to be generated when the  $L_i-L_j$ magnetic helicity 
is generated from a symmetric initial state. 
Moreover, if the initial condition already contains a non-zero 
helical $L_i-L_j$ magnetic field, a lepton asymmetry in each flavor
is generated as the magnetic helicity decays. 

As discussed in the previous section, such neutrino asymmetries 
can account for the ACT and SPT results as well as the EMPRESS measurements. 
Nevertheless, constructing a concrete scenario that
not only generates the required asymmetry but also preserves it
until BBN is highly nontrivial. 
In what follows, we explore the possibility of generating lepton
asymmetry in two flavors -- $L_i$ and $L_j$ --  in the context of axion inflation, 
which in turn leads to the electron-neutrino asymmetry sufficient to explain
the aforementioned measurements. 

\subsection{Generation of the helicity}
Let us examine how a helical $L_i$-$L_j$ gauge field is generated in our scenario of axion inflation, following Refs.~\cite{Jimenez:2017cdr,Domcke:2018eki}. See also Refs.~\cite{Kamada:2022nyt}.  
We begin by writing down the
Lagrangian of the inflaton field (axion) $\phi$ and the $L_i$-$L_j$ gauge field, which is given by
\begin{align}
    \frac{\mathcal{L}}{\sqrt{-g}}= \frac{1}{2}  \partial_\mu\phi\partial^\mu\phi- V(\phi)  -\frac{1}{4}  X_{\mu\nu}X^{\mu\nu}+ \frac{g_{\phi XX}}{4} \phi X_{\mu\nu}\tilde{X}^{\mu\nu},  
\end{align}
where $V(\phi)$ is the inflaton potential, and $g_{\phi XX}$ is the dimensionful axion-photon coupling.  
The metric is taken to be conformal, $ds^2=a^2(t)(d\eta^2-d\vec{x}^2)$, with $a(t)$ being the scale factor, and the index, e.g., $\mu$ runs for $\eta,x,y,z$. 
Here the covariant definition of the field strength tensor and its dual is employed:
$X^{\mu\nu}=g^{\mu\rho}g^{\nu\sigma} X_{\rho \sigma}$ and $\tilde{X}^{\mu\nu}=\frac{1}{2\sqrt{-g}}\epsilon^{\mu\nu\rho\sigma}X_{\rho \sigma}$
with $\epsilon^{\mu\nu\rho\sigma}$ being the totally asymmetric Levi-Civita symbol, $\epsilon^{0123}=+1$. 

We consider a setup in which inflation is driven by the potential energy
of a slowly rolling inflaton $\phi$, while the gauge-field production induced by the axion–gauge coupling develops on top of this background dynamics.
For our purpose of examining the generation of the $L_i-L_j$ magnetic helicity, we do not specify the detailed form of the inflaton potential $V(\phi)$. 
One may have in a mind cosine-like potential as in conventional axion inflation; 
however, the detailed shape of axion potential depends sensitively on the underlying high-energy theory~\cite{Svrcek:2006yi,Silverstein:2008sg,McAllister:2008hb,Burgess:2014oma,Nomura:2017ehb}. 
Instead, we characterize the background inflationary dynamics  by the Hubble parameter during inflation, $H_\mathrm{inf}$, and inflaton velocity through the dimensionless parameter  $\xi \equiv -g_{\phi XX} {\dot \phi}/2 H_\mathrm{inf}$.  
Under the slow-roll approximation, both quantities can safely be taken as approximately constants. 
Once we specify an inflaton potential, 
the values of these parameters at the time when the CMB scale exit the horizon and those near the end of inflation are related. 
The former are constrained by CMB and other observations, and consequently the latter -- relevant for magnetogenesis -- are also constrained.
Here, however, we take an agnostic stance toward this relation. 
Rather, we adopt the viewpoint that he parameter region suitable for our scenario can serve as a motivation for  model building of axion inflation and for exploring its  ultraviolet completion. 

For the gauge field $X_{\mu}=(X_0,\vec{X})$,
we adopt the radiation gauge, defined by the conditions $X_0=0$, $\nabla\cdot \vec{X}=0$. This gauge is effective in our case. 
We then perform a mode expansion of the field in the circular polarization basis as follows, 
\begin{equation}
    \vec{X}(\eta, \vec{x})=\int \frac{d^3 k}{ (2\pi)^{3/2}} \sum_{\sigma=\pm} \left[ \vec{\epsilon}^{\, (\sigma)} (\vec{k}) a_{\vec{k}}^{(\sigma)} X_{\sigma} (\eta,\vec{k})  e^{i\vec{k}\cdot \vec{x}}  + \mathrm{h.c.} \right],
\end{equation}
where polarization vector $\vec{\epsilon}^{\, (\pm)}$ for a given momentum $\vec{k}$ which form an orthonormal basis in complex vector space perpendicular to $\vec{k}$ satisfying $\vec{k} \cdot \vec{\epsilon}^{\, (\pm)}(\vec{k}) =0,\ i \vec{k} \times \vec{\epsilon}^{\,(\pm)} (\vec{k}) =\pm  k \vec{\epsilon}^{\, (\pm)}(\vec{k})
$ and $\vec{\epsilon}^{\,  (\sigma)*} (\vec{k}) \cdot\vec{\epsilon}^{\, (\sigma')} (\vec{k})= \delta^{\sigma \sigma'}$ with $k=|\vec{k}|$. The quantization is implemented by requiring the creation and annihilation operators, $a_{\vec{k}}^{(\sigma)}$  and $a_{\vec{k}}^{(\sigma) \dagger}$, to satisfy the usual canonical commutation relations,
$[a^{(\sigma)}_{\vec{k}},a_{\vec{q}}^{(\sigma') \dagger}]= \delta^{\sigma \sigma'} \delta(\vec{k}-\vec{q})$.

In a simple gauged U(1) $L_i-L_j$ extension of the SM, the induced current from non-perturbative production of $L_i-L_j$  charged fermions must be taken into account. 
To focus on the dynamics of the gauge field amplification,
however, we first neglect this effect. 
We will discuss how crucial in our scenario and how to take into account it later. 
The resulting equation of motion for the mode function in the inflationary background becomes,
\begin{align}
0=[\partial_{\eta}^2 + k(k \mp 2  \xi  a H_{\rm inf }) ]X_{\pm}(\eta,k), \label{modeeq}
\end{align}
where $\xi\equiv - \frac{g_{\phi XX}   \dot{\phi}}{2 H_{\rm inf }} $ is the instability parameter and $H_{\rm inf}$ is the Hubble constant during inflation.
As noted previously, we take them as constants.

A key feature of the system is a polarization-dependent instability that exponentially amplifies one of the two helicity modes ($\sigma = \pm$). Specifically, the positive helicity mode ($\sigma=+$) becomes unstable for $\xi > 0$, while the negative helicity mode ($\sigma=-$) is unstable for $\xi < 0$. 
In each case, the opposite helicity mode stays stable, and hence 
the amplified gauge fields are helical.
This amplification occurs for long-wavelength modes that satisfy the condition $k/a < 2 |\xi| H_{\mathrm{inf}}$. 
In the constant Hubble parameter approximation,
the spacetime is described by a de~Sitter geometry, where the scale factor is $a(t)=e^{H_{\rm{inf}} t}$ and the corresponding conformal time is $\eta=-1/(aH_{\rm{inf}})$.
Under this approximation, the mode equation Eq.~\eqref{modeeq} can be solved analytically. 
By imposing the standard Bunch-Davis vacuum condition in the asymptotic past ( $\lim_{-k\eta \to \infty} X_{\sigma}(\eta,\vec{k})= e^{-ik\eta}/\sqrt{2k}$ ), the solution is given in terms of the Whittaker function $W_{\kappa,\mu}$~\cite{Jimenez:2017cdr,Domcke:2018eki}, 
\begin{equation}
\label{eq:whittaker_solution}
    X_{\sigma}(\eta,\vec{k})= \frac{e^{\sigma \pi \xi/2}}{\sqrt{2k}} W_{- i\sigma \xi, 1/2 }(2 ik \eta). 
\end{equation}
This solution confirms that for a non-zero $\xi$ the mode corresponding to $\sigma = \text{sign}(\xi)$ experiences an exponential amplification.

We impose the condition $\xi > 0$, which is a crucial requirement to generate the correct positive sign of the lepton asymmetry to account for the helium abundance suggested by the ACT and SPT (and also EMPRESS).
By substituting the amplified mode solution ($X_+$) into the mode expansion, we can compute the quantum vacuum expectation values, denoted by $\langle\cdot\rangle$.
In the regime of $\xi \gtrsim 3$, the results exhibit minimal sensitivity to the UV cut-off~\cite{Jimenez:2017cdr}. Applying a momentum cut-off at the edge of the instability band, $k_{\text{max}} = 2\xi a H_{\rm{inf}}$, yields the following expectation values:
\begin{gather}
    \langle \vec{E}_X^2 \rangle =\frac{1}{2a^4} \int \frac{d^3 \vec{k}}{(2\pi)^3} \ |\partial_\eta X_{+}|^2\simeq 2.6 \times 10^{-4} \frac{e^{2\pi \xi}}{\xi^3 }H_{\rm{inf}}^4, \label{eq:axion_EX}\\
    \langle \vec{B}_X^2\rangle =\frac{1}{2a^4} \int \frac{d^3 \vec{k}}{(2\pi)^3} k^2 \ | X_{+}|^2\simeq 3.0 \times 10^{-4} \frac{e^{2\pi \xi}}{\xi^5 }H_{\rm{inf}}^4, \label{eq:axion_BX}\\
    \langle \vec{E}_X\cdot \vec{B}_X \rangle = -\frac{1}{2a^4}\int \frac{d^3k}{(2\pi)^3} \ k \partial_\eta |X_{+}|^2 \simeq - 2.6 \times 10^{-4} \frac{e^{2\pi \xi}}{\xi^4 }H_{\rm{inf}}^4. \label{eq:axion_EB}
\end{gather}
Here the physical electric field $\vec{E}_X$ and magnetic field $\vec{B}_X$ are defined in terms of the vector potential $\vec{X}$ and conformal time $\eta$ as, 
\begin{equation}
    \vec{E}_X = - \frac{\partial_\eta \vec{X}}{a^2} , \quad \vec{B}_X = \frac{\vec{\nabla}\times \vec{X}}{a^2}.
\end{equation}

The dominant contribution to these integrals arises from superhorizon modes ($k\lesssim a H_{\rm{inf}}$). This implies that the produced gauge fields are coherent and nearly constant over the Hubble scale. Furthermore, the negative sign of $\langle \vec{E}_X\cdot \vec{B}_X \rangle$ indicates that the electric and magnetic fields are, on average, aligned in an anti-parallel configuration. For a detailed discussion on the renormalization of these fields in the context of axion inflation, we refer the reader to Ref.~\cite{Ballardini:2019rqh}. The production rate of the physical helicity, $\mathcal{H}_X$, is governed by the following equation, 
\begin{equation}
    \partial_\eta (a^3 \langle \mathcal{H}_X \rangle) = -2 \int d^3 x \, a^4 \langle \vec{E}_X \cdot \vec{B}_X \rangle. 
\end{equation}
It implies that a non-zero value for the term $\langle \vec{E}_X \cdot \vec{B}_X \rangle$ sources the generation of helicity. Specifically, it can be seen that a positive magnetic helicity is induced by the end of the inflationary epoch, as expected.

To quantify this, we define the magnetic helicity density, $h_X$, as the spatial average of the total helicity over the comoving volume $\mathbb{V}$. This is given by, 
\begin{equation}
    h_X \equiv \frac{\langle \mathcal{H}_X \rangle}{\mathbb{V}}, \quad \text{where} \quad \mathbb{V} = a^3 \int d^3 x.
\end{equation}
The volume $\mathbb{V}$ represents the physical volume of a spatial hypersurface in Friedmann--Lema\^itre--Robertson--Walker (FLRW) coordinates. At the end of inflation, the magnetic helicity density is roughly estimated as, 
\begin{equation}
    h_X \sim - \left. \frac{2 \langle \vec{E}_X \cdot \vec{B}_X \rangle}{3H} \right|_{\mathrm{end}} \simeq 1.7 \times 10^{-4} \frac{e^{2 \pi \xi}}{\xi^4} H_{\rm{inf}}^3,
    \label{helinfend}
\end{equation}
which can be sizable if $\xi$ is sufficiently large. 
According to Eq.~\eqref{eq:csconservation}, one may expect that 
a large lepton asymmetry can be induced in the Universe without
inducing large baryon asymmetry if the magnetic helicity decays
after the EWPT.

\subsection{Fermion production and backreaction on the inflation dynamics}

In a simple gauged U(1)$_{L_i-L_j}$ extension of the SM, however, 
the fermions charged under $L_i-L_j$ remain effectively massless during inflation
and are therefore produced asymmetrically through the Schwinger effect. 
As a result, the total charge (the sum of the magnetic helicity and lepton asymmetry) vanishes at the end of inflation, 
\begin{equation}
    Q_L-\frac{g_{L_i-L_j}^2}{4 \pi^2} \langle \mathcal{H}_X \rangle =0. \label{eq:totalchargecons}
\end{equation}
In this case, the post-inflationary  coevolution
of the magnetic helicity and lepton asymmetry must be taken into account in order to determine whether a sizable electron–neutrino asymmetry can survive until BBN and thereby suppress the helium abundance,
in close analogy to the baryogenesis scenario discussed in Refs.~\cite{Domcke:2019mnd,Fukuda:2024pkh}. 
From this motivation, let us first discuss 
the fermion production or the Schwinger effect during inflation, 
which has been omitted in the discussion in the above. We consider 
a Dirac fermion $\psi$ interacting with a gauge field 
via the coupling $-g_{L_i-L_j} q_\psi \bar{\psi} \gamma^\mu X_\mu\psi$, where $q_\psi$ is the fermion's $L_i-L_j$ charge. Amplified gauge fields produce fermions and induce a fermionic current, which in turn backreacts on the inflation and gauge field dynamics. The generalization to multiple fermion species is straightforward by summing over their individual contributions.

Ideally, one would evaluate the fermion production 
by solving simultaneously the Dirac equation and the gauge-field equation of motion. 
However, no closed-form analytic solution is known, and 
even a reliable numerical scheme has yet to be developed. 
We therefore resort to an analytic treatment based on controlled approximations. 
Specifically, we first compute the fermion production in a fixed background of  the amplified gauge field, and subsequently feed this back into 
the gauge-field equation to assess the backreaction. 
Even the fermion production in a background gauge field, {\it i.e.},
the Schwinger effect~\cite{Schwinger:1951nm}, 
is not analytically solvable in general~\cite{Gelis:2015kya}. 
In the present setup, the gauge fields behave
as stochastic variables~\cite{Fujita:2022fit}
whose power spectrum is described by the Whittaker functions, 
which again is also not analytically solvable. However, as argued above, 
the dominant contribution come from the superhorizon modes, for which the gauge field can be approximated by constant, anti-parallel electric and magnetic fields. In this limit, the fermion production becomes analytically tractable and admits a transparent physical interpretation. 

In a constant magnetic-field background, the fermion spectrum is quantized into Landau levels. The lowest Landau level carries a definite chirality, while all higher levels are chirally symmetric. When an electric field is  
applied parallel to magnetic field, fermionic states are driven along these Landau levels, inducing a fermion current~\cite{Nielsen:1983rb,Fukushima:2008xe}.
This induced current receives two distinct contributions:  the lowest Landau level (LLL) component, 
which encodes the chiral anomaly~\cite{Nielsen:1983rb}, and the contributions from higher Landau levels (HLLs), which correspond to the Schwinger effect~\cite{Heisenberg:1936nmg,Schwinger:1951nm}, namely, tunneling between discrete energy levels. 
For a background  $\vec{E}_X=(0,0,E_X)$ and $\vec{B}_X=(0,0,-B_X)$, the resulting current is given by~\cite{Domcke:2018eki}
\begin{equation}
g_{L_i-L_j}q_\psi \langle J^z_\psi \rangle = \frac{(g_{L_i-L_j} |q_\psi| )^3}{6 \pi^2} \coth \left(\frac{\pi B_X}{E_X} \right) E_X B_X \frac{1}{H_{\rm inf }}, \label{eq:Jpsi}
\end{equation}
where we have neglected particle scattering effects.
In the case with multiple charged fermions, we just sum up all the contributions to evaluate the current. 
In the SM, each flavor contains one left-handed SU(2) doublet fermion and one right-handed SU(2) singlet fermion, both carrying
an absolute $L_i-L_j$ charge of 1. 
Accordingly, Eq.~\eqref{eq:Jpsi} should be modified by the replacing,
\begin{equation}
    |q_\psi|^3 \rightarrow 2 \times \frac{1}{2} \times (2+1) = 3, 
\end{equation}
where the  factor of 2 accounts for
2 lepton flavors, the factor of 1/2 converts the contribution from a Dirac fermion used in Eq.~\eqref{eq:Jpsi} to that for a Weyl fermion, 
and the sum 2+1 reflects the number of degree of freedom contributing to the current per flavor. 
If right-handed neutrinos are present, they add 
one more Weyl fermion per flavor, leading to  $|q_\psi|^3 \rightarrow 4$. 

One might consider solving Maxwell’s equations with the induced current Eq.~\eqref{eq:Jpsi} to evaluate the back reaction from the particle production. 
However, this remains challenging, as it requires computing the time-dependent averages of the electric and magnetic fields at each moment.
Therefore, additional approximations are necessary in order to estimate the impact of particle production on the gauge-field dynamics.
While several approaches exist~\cite{Domcke:2018eki,Gorbar:2021rlt,Gorbar:2021zlr,Fujita:2022fwc}, a quantitatively precise consensus remains elusive. Here we adopt the ``equilibrium estimate''~\cite{Domcke:2018eki} to provide a concrete illustration.

The idea of this estimate is as follows. 
Starting from Maxwell's equations with the induced current, one can derive the energy transfer equation for the gauge field, 
\begin{equation}
\dot{\rho}_X=-4 H_{\rm{inf}} \rho_X - 2 \xi H_{\rm inf} \langle \vec{E}_X\cdot \vec{B}_X \rangle - g_{L_i-L_j} q_\psi \langle \vec{E}_X\cdot \vec{J}_\psi\rangle.  \label{eq:EOMrhoX}
\end{equation}
Here, $\rho_X$ is the expectation value of the gauge field energy density, 
\begin{equation}
    \rho_X= (1/2 \mathbb{V}) \int d^3 x \langle \vec{E}_X^2 + \vec{B}_X^2 \rangle, \label{eq:enegaugefield}
\end{equation} and $\vec{J}_\psi$ is the induced fermion current.
In the absence of induced currents,
energy transfer from the axion to the gauge field is solely balanced by 
cosmic expansion, resulting into a stationary electromagnetic field configuration. 
This explains why the estimates in Eqs.~\eqref{eq:axion_EX}-\eqref{eq:axion_EB} are time-independent. When charged fermions are present, the induced current introduces an additional channel of energy transfer
--  from the gauge field to the fermion sector --  as reflected in the last term of Eq.~\eqref{eq:EOMrhoX}. 
Assuming that the system reaches a dynamical equilibrium with $\dot{\rho}_X=0$, where energy injection from axion dynamics balances both fermion production and cosmic expansion, 
Eq.~\eqref{eq:EOMrhoX} can be approximated as
\begin{equation}
0=-2 H_{\rm inf } \langle \vec{E}_X^2+\vec{B}_X^2 \rangle -2 \xi H_{\rm inf} \langle \vec{E}_X\cdot \vec{B}_X \rangle - g_{L_i-L_j} q_\psi \langle \vec{E}_X\cdot \vec{J}_\psi\rangle.
\label{eq:backreaction_equilibrium}
\end{equation}
Further approximating $\langle \vec{E}_X^2 \rangle \simeq E_X^2$, $\langle \vec{B}_X^2 \rangle \simeq B_X^2$, $\langle \vec{E}_X\cdot \vec{B}_X \rangle \simeq -E_X B_X$, and $\langle \vec{E}_X\cdot \vec{J}_\psi\rangle \simeq E_X \langle J^z_\psi \rangle$, and substituting the induced current from Eq.~\eqref{eq:Jpsi}, we obtain
\begin{equation}
0=-2 H_{\rm inf } (E_X^2+B_X^2)+2\xi_{\rm eff} H_{\rm inf } E_X B_X, \label{eq:equilibriumxieff}
\end{equation}
where the effective instability parameter is defined as
\begin{equation}
\xi_{\rm eff} \equiv \xi-\frac{(g_{L_i-L_j} |q_\psi| )^3}{12 \pi^2} \coth \left(\frac{\pi B_X}{E_X} \right) \frac{E_X }{ H_{\rm inf }^2}. \label{eq:xieff}
\end{equation}
This clearly demonstrates that fermion production acts to suppress helicity generation in the gauge field.

To estimate the resulting gauge field strength, we assume a quasi-stationary electromagnetic configuration, with $\xi_{\rm eff}$ being treated as constant, as will be discussed in the following subsection. If $\xi_{\rm eff}$ is regarded as an independent constant parameter, the equation of motion for $\vec{X}$ with the fermion induced current becomes formally identical to the fermion-free case in Eq.~\eqref{modeeq}, 
with $\xi$ replaced by $\xi_{\rm eff}$. Consequently, the solutions for $E_X$ and $B_X$ take the same form as the analytical solutions in Eqs.~\eqref{eq:axion_EX} and \eqref{eq:axion_BX}, with the replacement $\xi \to \xi_{\rm eff}$.
By solving Eq.~\eqref{eq:xieff} with these expressions of $E_X$ and $B_X$ substituted, $\xi_\mathrm{eff}$ can be determined as a function of $\xi$.
It has been turned out that $\xi_\mathrm{eff}$ is no longer 
linear in $\xi$, but monotonically increasing function of $\xi$.
The magnetic helicity at the end of inflation is then  evaluated using Eq.~\eqref{helinfend}, again with $\xi$ replaced by $\xi_\mathrm{eff}$. 
As previously discussed, the produced fermions carry lepton asymmetries
such that the sum of the magnetic helicity and the lepton asymmetry vanishes (Eq.~\eqref{eq:totalchargecons}).
We evaluate the lepton number
density at the end of inflation as
\begin{equation}
    n_{L} \equiv \frac{Q_{L}}{\mathbb{V}}= \frac{g_{L_i-L_j}^2}{4 \pi^2}  
    h_X \simeq 4.4 \times 10^{-6} 
    \frac{g_{L_i-L_j}^2e^{2 \pi \xi_\mathrm{eff}}}{\xi^4_\mathrm{eff}} H_{\rm inf}^3. 
\end{equation}

\subsection{Upper bound on the asymmetry}
According to the previous argument, one might expect that an arbitrarily large lepton asymmetry could be obtained by increasing $\xi$. 
However, the energy densities of the produced gauge fields and fermions 
must not exceed that of the inflaton. 
If $\xi$ (or $\xi_\mathrm{eff}$)  becomes so large that this condition is violated, the backreaction on the inflaton dynamics can no longer be neglected. 
Here we conservatively require that the sum of
the energy density of  U(1)$_{L_i-L_j}$ gauge field ($\rho_X$) 
and  of the fermions ($\rho_\psi$) remain below 
$ 1\% $ of the inflaton energy density. 
This bound can be expressed as
\begin{align}
    \rho_X+\rho_{\psi} < 0.01 \cdot \rho_\phi =0.01 \cdot 3 H_\mathrm{inf}^2 M_\mathrm{pl}^2, \label{eq:negbr}
\end{align}
where $M_\mathrm{pl}$ is the reduced Planck mass. 
While one might in principle perform lattice simulations including such strong backreaction on inflaton dynamics from the energy density from the produced particles~\cite{Garcia-Bellido:2023ser,Figueroa:2023oxc,vonEckardstein:2023gwk,Figueroa:2024rkr}, 
simulations that incorporate the  Schwinger effect have not yet been explored. 

The energy density of the U(1)$_{L_i-L_j}$ gauge field can be directly evaluated by Eqs.~\eqref{eq:axion_EX}, \eqref{eq:axion_BX}, and \eqref{eq:enegaugefield}, 
\begin{equation}
    \rho_X = \frac{1}{2} \left( \langle \vec{E}_X^2\rangle + \langle \vec{B}_X^2\rangle\right)  \simeq \frac{1}{2} \langle \vec{E}_X^2\rangle \simeq 1.3 \times 10^{-4} \frac{e^{2\pi \xi_\mathrm{eff}}}{\xi^3_\mathrm{eff} }H_{\rm{inf}}^4 \left(\sim H_\mathrm{inf} h_X \right), 
\end{equation}
where $\xi$ in Eqs.~\eqref{eq:axion_EX} is replaced by $\xi_\mathrm{eff}$, and the magnetic contribution is omitted since it is subdominant. 
In contrast, the energy density of the produced fermions requires a more careful treatment, because one must  take into account the acceleration and possible thermalization at each Landau level. 
In the following, we compute $\rho_\psi$
following the method developed in Ref.~\cite{Domcke:2018eki}. 

To evaluate the energy density of fermions produced by the strong electromagnetic field, 
we consider a configuration in which the electric and magnetic fields are both aligned along the $z$-direction, with the magnitudes $E$ and $B$, respectively.
Here the electric field is assumed to be turned on at only during the finite interval $0<t<\tau$. 
By quantizing the Weyl fermion in terms of Landau levels and solving the Weyl equation in this background,  one obtains the evolution equation for 
the energy density of the $n$-th Landau level as  
\begin{align}
\label{eq:my_rho_dot_n}
\dot{\rho}_{R}^{n} &= \frac{1}{\tau}\frac{g_{L_i-L_j}|Q|B}{2\pi}\int \frac{dp_{z}}{2\pi} \, p_{z} \, \theta(-p_{z})\theta(p_{z}+gQE\tau) \exp\left[-\frac{2\pi nB}{E}\right] \nonumber \\
&=\frac{(g_{L_i-L_j}|Q|)^3}{8\pi^2} E^2 B\exp \left[ -\frac{2\pi nB}{E}\right] \tau, 
\end{align}
for the right-handed fermion where $Q$ is its $L_i-L_j$ charge. Here for a moment the cosmic expansion is neglected. 
Summing over all Landau levels with $n \geq 1$,  
the evolution equation for the total energy density of the HLL is obtained as
\begin{align}
    \dot{\rho}_R^{\text{HLL}} 
    &=\frac{(g_{L_i-L_j}|Q|)^3}{8\pi^2}E^2 B \tau \frac{1}{ \text{exp}[\frac{2\pi B}{E}]-1}. 
\end{align}

Before performing the time integration to evaluate the energy density of the HLL fermions, we  
clarify an  
assumption of the calculation. 
In a dynamically expanding de Sitter universe, the physical background fields $E$, $B$, and the Hubble parameter 
do evolve
with time.
However, it can be evaluated that 
the characteristic timescale for Schwinger fermion production is
much shorter  
than the Hubble time. 
Therefore, when analyzing the microphysical dynamics of particle creation, it is justified 
to treat $E$, and $B$  
as quasi-static, effectively constant background quantities, and the effect of cosmic expansion is 
neglected. 
This ensures the consistency of our treatment in which the electromagnetic fields are considered time-independent during the particle production interval.

We then turn on the cosmic expansion adiabatically to evaluate the energy density of the produced fermions, 
by replacing $E$ and $B$ with their comoving counterparts. 
Performing the time integral, we obtain 
\begin{align}
    \rho^{\text{HLL}}_R&=\frac{(g|Q|)^3}{32
    \pi^2}\frac{E^2 B}{H_\mathrm{inf}^2} \frac{1}{ \text{exp}[\frac{2\pi B}{E}]-1}=\rho_L^{\text{HLL}}=\bar{\rho}_R^{\text{HLL}}=\bar{\rho}_L^{\text{HLL}}.  
\end{align}
Here, we note that contribution also come from antiparticles as well as the left handed particles, which are the same to the right-handed particles.
Thus the total
energy density of the HLL fermions is given by
\begin{align}
    \rho^{\text{HLL}}_\psi=\frac{(g|Q|)^3}{8\pi^2}\frac{E^2 B}{H_\mathrm{inf}^2} \frac{1}{ \exp[\frac{2\pi B}{E}]-1}. 
\end{align}
Note that in Eq.~\eqref{eq:my_rho_dot_n} we have neglected the transverse momentum, which becomes sizable 
for large $n$. 
However, the contributions from such levels are exponentially 
suppressed, therefore the resulting estimate remains essentially unchanged even when they are included.

If one were to naively perform the same analysis
for the LLL, 
its contribution to the energy density would be estimated 
as $\rho_\psi^\mathrm{LLL}=\frac{(g|Q|)^3}{16 \pi^2}\frac{E^2 B}{H_\mathrm{inf}^2}$, such that the total energy density would be given by 
$\rho_{\psi}=\frac{(g|Q|)^3}{16
\pi^2}\frac{E^2 B}{H_\mathrm{inf}^2} \coth[\pi \frac{B}{E}]$.
However, as argued in Ref.~\cite{Domcke:2018eki}, 
{this naive estimate fails because one must take into account scattering among the produced fermions and the resulting thermalization.
When the scattering rate exceeds the Hubble time, the produced fermions thermalize. 

The scattering rate for the LLL fermions are evaluated as follows. 
For fermions with number density $n_\psi$ and center-of-mass energy $s_\mathrm{cm}$, it is given by
\begin{equation}
    \Gamma_\mathrm{sc}=n_\psi \sigma_\mathrm{sc}, \quad \text{with} \quad \sigma_\mathrm{sc} = \frac{4 \pi \alpha^2}{3s_\mathrm{cm}}, 
\end{equation}
where $\alpha$ is the fine structure constant associated with the strongest gauge interaction that the fermion charges. 
Assuming that the LLL number density is determined by the balance between production and dilution due to cosmic expansion, 
we estimate $n_\psi^\mathrm{LLL} = {\dot n}_\psi^\mathrm{LLL} H^{-1}_\mathrm{inf} = (g_{L_i-L_j}^2 Q^2/2 \pi^2) EBH^{-1}$.  
The center of mass energy is evaluated as time-dependent quantity, $s_\mathrm{cm}=2 (g_{L_i-L_j} Q E \tau)^2$. Imposing the condition $\Gamma_\mathrm{sc} = \tau^{-1}$ determines the scattering time 
\begin{equation}
    \tau_\mathrm{sc} = 1.1 \times 10^{-5} \left(\frac{\alpha}{10^{-2}}\right)^2\frac{B}{E} H^{-1}_\mathrm{inf}. 
\end{equation}
Since the electric field is typically stronger than  magnetic field, we conclude that LLL fermions thermalize. 
Because their acceleration stops at the scattering time, 
the energy density of the LLL fermions can be estimated as 
\begin{equation}
    \rho_\psi^\mathrm{LLL} \simeq n_\psi^\mathrm{LLL} \sqrt{\frac{s}{2}} \simeq 5.4 \times 10^{-7} \left(\frac{\alpha}{10^{-2}}\right)^2 (g|Q|)^3 \frac{EB^2}{H_\mathrm{inf}^2}. 
\end{equation}
A similar analysis shows that the HLL fermions do not thermalize. 
Therefore, the total energy density of the produced fermions can be approximated by the HLL contribution, 
\begin{align}
\rho_{\psi}=\rho^{\text{LLL}}_\psi+\rho^{\text{HLL}}_\psi \simeq \rho^{\text{HLL}}_\psi =\frac{(g|Q|)^3}{8\pi^2}\frac{E^2 B}{H^2} \frac{1}{ \exp[\frac{2\pi B}{E}]-1} \simeq \frac{(g|Q|)^3}{16\pi^3}\frac{E^3}{H^2}. \label{eq:rhopsi}
\end{align}

Now we are ready to constrain the parameter space using Eq.~\eqref{eq:negbr}. 
For convenience, we introduce the helicity-to-entropy ratio, normalized so as to directly estimate the generated lepton asymmetry, 
\begin{equation}
    \eta_\mathrm{H} \equiv \frac{g_{L_i-L_j}^2}{8\pi^2} \frac{h_X}{s},   \label{eq:etahdef}
\end{equation}
such that the parameter regions for $n_{L_e}/s$ suggested by ACT (Eq.~\eqref{eq:nueasymACT}) and SPT (Eq.~\eqref{eq:nueasymSPT}) are directly mapped onto
those for $\eta_\mathrm{H}$; see Eq.~\eqref{eq:csconservation}. 
Assuming 
instant reheating, the entropy density at the end of inflation is evaluated as 
\begin{align}
    s=\frac{2 \pi^2}{45} g_{* s} T_\mathrm{re}^3=\frac{2 \pi^2}{45} g_{* s}\left(\frac{90}{\pi^2 g_*}\right)^{3 / 4}\left(H_{\mathrm{inf}} M_{\mathrm{pl}}\right)^{3 / 2},   
\end{align}
where $T_\mathrm{re}$ denotes the reheating temperature. 
The helicity generated during axion inflation (Eq.~\eqref{helinfend}), 
including the Schwinger effect is given by
\begin{align}
    h_X=1.7 \times 10^{-4} \frac{e^{2 \pi \xi_{\mathrm{eff}}}}{\xi_{\mathrm{eff}}^4} H_{\mathrm{inf}}^3. 
\end{align}
Combining these expressions, we obtain
\begin{align}
    \eta_\mathrm{H}= 9.4 \times 10^{-7} g_*^{-1 / 4} g_{L_i-L_j}^2 \frac{e^{2 \pi \xi_{\mathrm{eff}}}}{\xi_{\mathrm{eff}}^4}\left(\frac{H_{\mathrm{inf}}}{M_{\mathrm{pl}}}\right)^{3 / 2},  \label{eq:etaH}
\end{align}
where we have approximated $g_{* s}=g_*$. 

\begin{figure}
    \centering
    \begin{subfigure}[b]{0.48\textwidth}
    \centering
        \includegraphics[width=\textwidth]{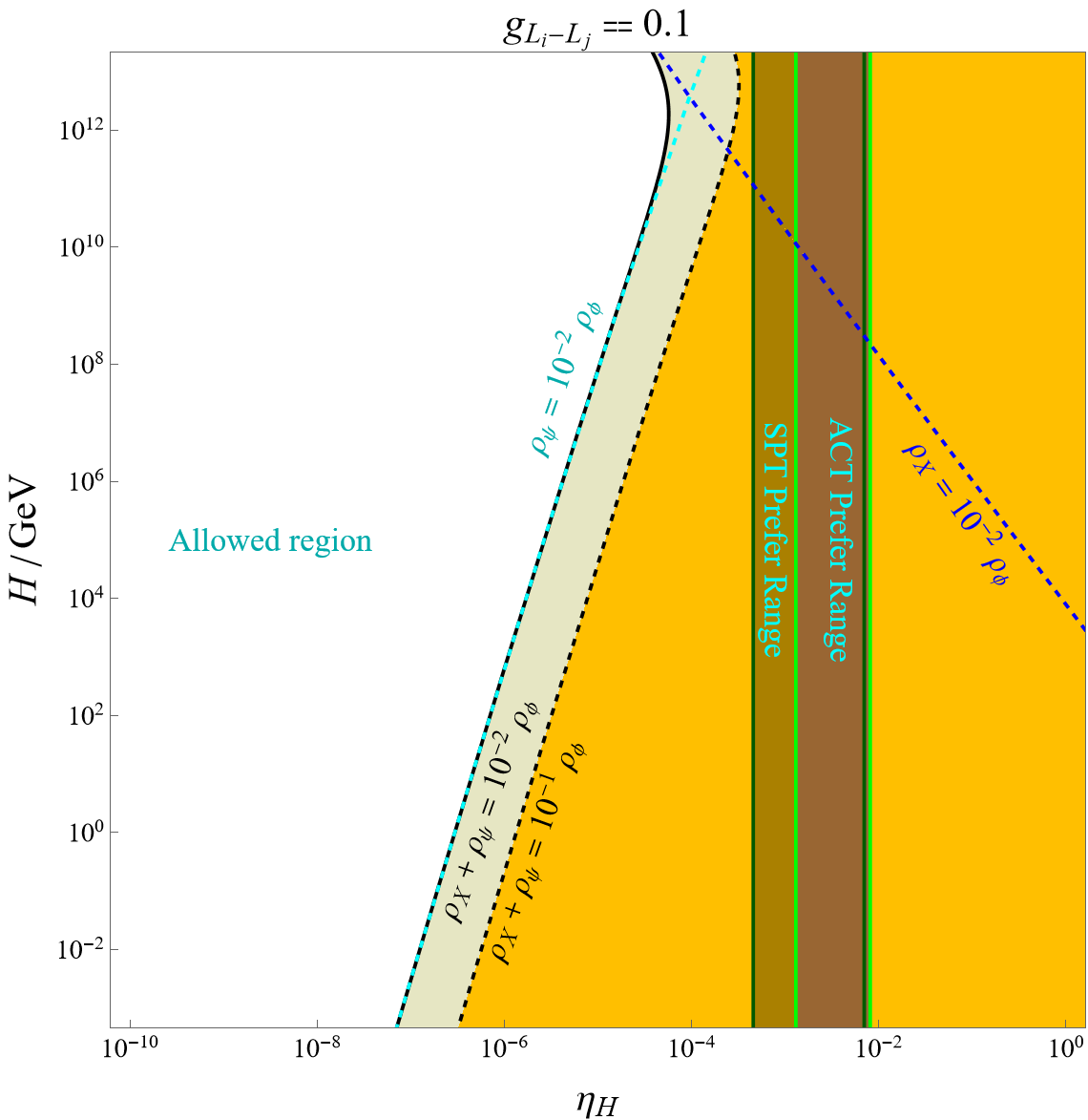}
    \end{subfigure}
    \hfill 
    \begin{subfigure}[b]{0.48\textwidth}
        \centering
        \includegraphics[width=\textwidth]{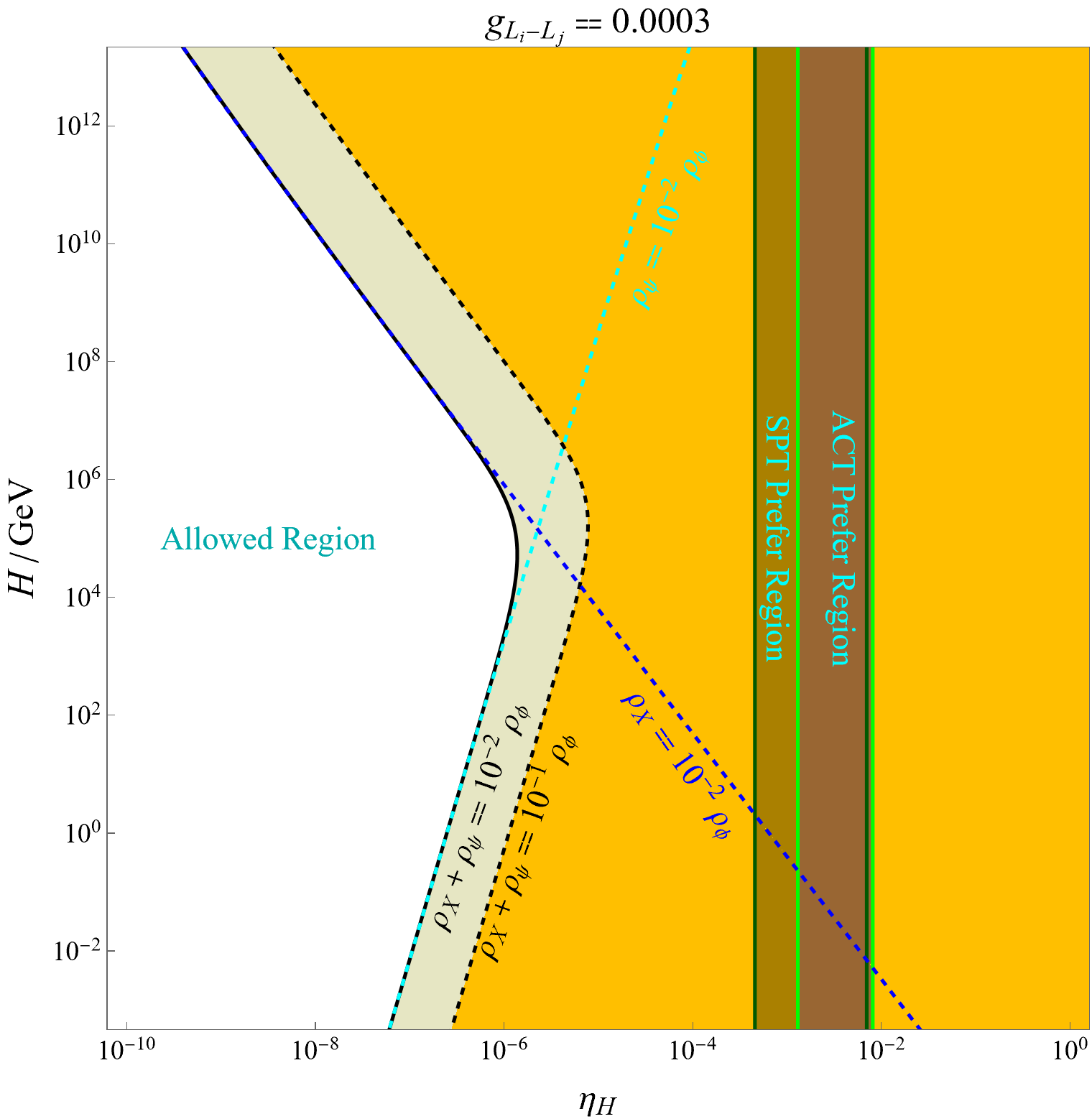}
    \end{subfigure}
    {\caption{
    \justifying
    The parameter space excluded by non-negligible backreaction from the energy densities of the produced gauge fields and fermions is shown in the plane of the Hubble parameter during inflation and the helicity-to-entropy ratio, for $g_{L_i-L_j}= 0.1$ (left) and 0.0003 (right). 
    The black solid line correspond to $\rho_{\psi}+\rho_{X}=0.01\rho_{\phi}$, while the black dashed line indicate $\rho_{\psi}+\rho_{X}=0.1 \rho_\phi$  for reference.
    The yellow-shaded regions to the right of these lines are excluded. 
    The thick and thin blue dashed lines  
    indicate the conditions $\rho_{X}=0.01\rho_{\phi}$ and $\rho_{\psi}=0.01\rho_{\phi}$, respevtively, showing which contribution dominates the backreaction. 
    The thick and thin brown shaded regions denote the 
    parameter space preferred by the ACT and SPT results, assuming that the corresponding electron neutrino asymmetry is generated~(Eq.~\eqref{eq:nueasymACT}). Note that the upper bounds of the ACT and SPT results are almost the same. 
    }
    \label{withschwinger}}
\end{figure}
Using Eq.~\eqref{eq:etaH}, both $\rho_X$ and $\rho_\psi$ can be expressed  as functions of $\eta_\mathrm{H}$ and $H_\mathrm{inf}$. 
Figure~\ref{withschwinger} shows the constraints on the parameter space satisfying Eq.~\eqref{eq:negbr} in the $\eta_\mathrm{H}-H_\mathrm{inf}$ plane
for $Q=3$ and $g_{L_i-L_j}=0.1$ and 0.0003. 
For reference, we also show the parameter region  satisfying $\rho_\phi+\rho_\psi < 0.1 \rho_\phi$. 
While the condition $\rho_X < 0.01 \rho_\phi$ determines the constraint at higher $H_\mathrm{inf}$, the bound at lower $H_\mathrm{inf}$ is  set by $\rho_\psi < 0.01\rho_\phi$. 
The latter constraint is almost independent of $g_{L_i-L_j}$ because, using Eqs.~\eqref{eq:axion_BX}, \eqref{eq:axion_EX}, \eqref{eq:rhopsi} and  \eqref{eq:etaH}, one finds $\rho_\psi \propto \eta_\mathrm{H}^{3/2} H_\mathrm{inf}^{7/4}$, which does not depend on $g_{L_i-L_j}$. 
From the kinks of the constraint curves, we obtain the upper bound on the helicity-to-entropy ratio, 
\begin{equation}
    \eta_\mathrm{H} < 3.2 \times 10^{-6} \left(\frac{g_{L_i-L_j}}{10^{-3}} \right)^{0.69} \left( \frac{0.01}{(\rho_X+\rho_\psi)/\rho_\phi}\right)^{0.75}
\end{equation}
This implies that 
the resultant asymmetry cannot reach the level required to explain smaller helium abundance suggested by ACT, SPT and EMPRESS (Eq.~\eqref{eq:nueasymACT}), 
even if one invokes mechanisms to transfer this asymmetry into an electron neutrino asymmetry by the time of BBN; 
see, {\it e.g.}, Ref.~\cite{Fukuda:2024pkh}. 
One may wonder whether this difficulty can be alleviated by  considering a larger gauge coupling or allowing more gauge fields and fermion energy density, up to $\rho_X+\rho_\psi \sim \rho_\phi$. However, the ACT-preferred regime can be reached only for a large Hubble parameter, in which case  baryon overproduction is difficult to avoid.  
The difficulty originates from the Schwinger particle production. 
In the next section, we therefore examine the possibility of suppressing the Schwinger effect during inflation,
so that the $L_i-L_j$ magnetogenesis  can lead to a sufficiently large lepton asymmetry capable of 
accounting for the lower helium abundance suggested by ACT, SPT and EMPRESS.

\section{A model without Schwinger effect \label{sec:MGwithoutSE}}

We now investigate whether  a large electron neutrino 
asymmetry can be generated by the time of BBN within axion inflation coupled to a 
U(1)$_{L_i-L_j}$ gauge field. 
The main obstacles idetified in the previous section are: 
\begin{itemize}
    \item the excessive fermion energy density induced by the Schwinger effect, and 
    \item the overproduction of baryon asymmetry through sphaleron processes once a large lepton asymmetry is generated. 
\end{itemize}
These considerations suggest that a sizable electron neutrino asymmetry at BBN may still be achievable  
if  the following conditions are realized:  
\begin{enumerate}
    \item During inflation, the $L_i-L_j$ helicity is generated while Schwinger effect is effectively switched off,
    \item After reheating, the asymmetry remains stored in the magnetic helicity for some time, 
    \item After the EWPT, the magnetic helicity decays,  transferring its stored asymmetry into lepton asymmetry before BBN. 
\end{enumerate}
In the following, we examine the setup and conditions under which  this scenario 
can be realized.

\subsection{Switching off the Schwinger effect}

Let us first examine how the Schwinger effect can be switched off during inflation. 
This can be achieved if U(1)$_{L_i-L_j}$-charged leptons become sufficiently heavy. 
As a concrete working example, let us consider the following interaction, 
\begin{equation}
    {\cal L}_\mathrm{int} \ni - \frac{1}{2}(m_S^2 -c_S R) S^2 - \frac{\lambda_S}{4}S^4 - (m_H^2 - \lambda_{SH} S^2) |\mathcal{H}|^2 -  \lambda_H |\mathcal{H}|^4 +y(S) L\mathcal{H}e_R^i + f(S) \frac{(L^i \mathcal{H})^2}{\Lambda_W}.  
\end{equation}
Here we introduced a spectator scalar field $S$ with a non-minimal
coupling to gravity, which 
also interacts with the SM Higgs $\mathcal{H}$. 
The parameters are defined as follows. 
$m_S$ is the mass of $S$, $c_S$ is the non-minimal coupling coefficient, 
$\lambda_S$ is its quartic coupling.  
$m_H^2 \simeq -(125 \mathrm{GeV})^2$ is the Higgs mass, 
$\lambda_{SH}$ is the  Higgs-$S$ portal coupling, and
$\lambda_H\simeq 0.1$ is the Higgs quartic coupling. 
$y(S)$ and $f(S)$ provide the $S$-dependent Yukawa coupling and the Weinebrg operator, respectively, 
both of which can be up to order of the unity. 
$\Lambda_W$ denotes the energy scale suppressing the Weinberg operator. 

During inflation, the Ricci scalar is well approximated as $R \simeq 12H^2_\mathrm{inf}$ under the slow-roll condition. Consequently, the effective potential for $S$ is given by
\begin{align}
V_S(S) & =\frac{1}{2}m_S^2S^2-6 c_S H^2_\mathrm{inf} S^2+\frac{1}{4} \lambda_S S^4.
\end{align}
For the $S$ field to acquire a non-zero 
expectation value, 
its effective mass squared  must be negative. 
We therefore require that the curvature-induced mass term dominates over the bare mass, 
\begin{align}
c_S > \frac{m_S^2}{12H^2_\mathrm{inf}}, \label{eq:sdestab}
\end{align}
so that the potential can be simplified to
\begin{align}
V_S(S) \simeq -6 c_S H^2_\mathrm{inf} S^2+\frac{1}{4} \lambda_S S^4.
\end{align}
The resulting negative mass term then drives the $S$ field 
to acquire a non-zero expectation value, 
\begin{align}
S_0 = \sqrt{\frac{12c_S H^2_\mathrm{inf} }{\lambda_S}}.
\end{align}
For $c_S \gg 1/24$, the mass of $S$ around the 
potential minimum is heavier than the Hubble parameter during inflation,and the field $S$ is sufficiently stabilized. 

Once the $S$ field acquires a nonzero expectation value,
it can be treated as a classical background field. 
The effective potential for the Higgs field then takes the form
\begin{align}
V_H(\mathcal{H}) =  (m_H^2 - \lambda_{SH} S_0^2) |\mathcal{H}|^2 +  \lambda_H |\mathcal{H}|^4.
\end{align}
If the $S$-induced mass term dominates over the bare Higgs mass term, $\lambda_{SH} S_0^2 \gg |m_H^2|$, which can be easily satisfied
in the present setup, 
we can neglect the first term, and the Higgs potential is well approximated by
\begin{align}
V_H(\mathcal{H}) =  -\lambda_{SH} S^2_0 |\mathcal{H}|^2 +  \lambda_H |\mathcal{H}|^4. 
\end{align}
As a result, the Higgs field acquires an induced expectation value,
\begin{align}
|\mathcal{H}_0| = \sqrt{\frac{\lambda_{SH}}{2\lambda_H}}S_0.
\end{align}

With this Higgs expectation value, the $S$-dependent Yukawa interaction and the Weinberg operator generated masses for the charged leptons and neutrinos, 
\begin{equation}
    m_{e_i} = y(S_0) \mathcal{H}_0 \simeq \sqrt{\frac{6 c_S \lambda_{SH}}{\lambda_S \lambda_H}} H_\mathrm{inf}, \quad m_{\nu_i}= f(S_0) \frac{\mathcal{H}_0^2}{\Lambda_W} \simeq \frac{6 c_S \lambda_{SH}}{\lambda_S \lambda_H} \frac{H_\mathrm{inf}^2}{\Lambda_W}, 
\end{equation}
where, for simplicity,  we have set $y(S_0)=f(S_0)= 1$. 
In order to efficiently suppress the Schwinger effect, the fermion masses are required to satisfy the condition
\begin{equation}
m_{e_i},\, m_{\nu_i} \gtrsim \sqrt{g_{L_i-L_j} E_X/\pi}. \label{eq:heavylepton}
\end{equation}

Let us now examine the consistency conditions for this setup to be viable. 
First, the $S$ field must remain stabilized at its expectation value $S_0$ 
in the presence of the induced Higgs expectation value $\mathcal{H}_0$. 
This requirement implies that the curvature induced mass of $S$ dominates over the Higgs-induced contributions, which can be  expressed as\footnote{For $\lambda_S \lambda_H <\lambda_{SH}^2 < 4 \lambda_S \lambda_H$, the potential is unbounded from the below, but the tunneling rate during inflation as well as at zero temperature is sufficiently suppressed as long as $\lambda_H$ is not so large.}
\begin{equation}
    24 c_s H_\mathrm{inf}^2 > \lambda_{SH} \mathcal{H}_0^2 \Rightarrow \lambda_{SH}^2 < 4 \lambda_S \lambda_H. \label{eq:higgsexp}
\end{equation}
Second, the induced potential energy of the Higgs and the  $S$ field must not spoil the inflationary dynamics. 
The energy density contribution from the $S$ field is given by 
\begin{align}
\rho_S = V_S(S_0) = - 36\frac{c_S^2}{\lambda_S}H^4_\mathrm{inf}, 
\end{align}
and that for contribution from the Higgs field is estimated as
\begin{align}
\rho_H = V_{H}(H_0) = -36\frac{\lambda_{SH}^2}{\lambda_H}\frac{c_S^2}{\lambda_S^2}H^4_\mathrm{inf}.
\end{align}
To ensure that these spectator fields do not 
significantly backreact on the inflationary background, 
their total energy density must remain subdominant compared to the inflaton energy density $\rho_\phi$, namely,
\begin{align}
|\rho_H + \rho_S| < \rho_\phi. \label{eq:noBR}
\end{align}
Third, the $S$ field must decay sufficiently rapidly after inflation so as to avoid inducing the early-matter dominated era. 
In our setup, without introducing additional interactions,  the $S$ field can 
decay through the $S$-dependent Yukawa coupling. 
Taking $f(S) = S/\Lambda_S$ with $\Lambda_S \simeq S_0$, 
the decay rate is estimated as $\Gamma \sim (m_S^3/\Lambda_S^2)/256 \pi^3 \simeq (\lambda_S m_S^3/c_S H_\mathrm{inf}^2)/3172 \pi^3$. 
Here we neglect the effect of Higgs expectation value,
since it typically oscillates around the origin on a time scale much shorter than the inverse of the decay rate. 
Requiring that the decay rate exceeds the Hubble parameter at the end of inflation
implies that $\lambda_S$ must be sufficiently large, 
\begin{equation}
    \lambda_S > 3172 \pi^3 c_S \left(\frac{H_\mathrm{inf}}{m_S}\right)^3. \label{eq:sdecay}
\end{equation}
This constraint, however, can be relaxed if 
additional interactions of the $S$ field are introduced. 
If the reheating temperature exceeds $m_S$, the $S$ field condensate may be immediately  ``melted'', in which case the above condition is no longer required.
Finally, 
to ensure the validity of the effective field theory (EFT), the expectation values of 
the $S$ and Higgs fields must remain below the cutoff scale
\begin{align}
S_0, H_0 < \Lambda_W.\label{eq:EFT}
\end{align}

In summary, the model considered here must satisfy the following six conditions to effectively switch off the Schwinger effect:
\begin{enumerate}
\item The $S$ field is destabilized from the origin (Eq.~\eqref{eq:sdestab}): $c_S > \dfrac{m_S^2}{12H^2}$;
\item Leptons become sufficiently heavy (Eq.~\eqref{eq:heavylepton}): $m_{e_i}, m_{\nu_i} > \sqrt{g_{L_i-L_j}E_X/\pi}$;
\item A non-zero expectation value of the $S$ field is stabilized (Eq.~\eqref{eq:higgsexp}): $\lambda_{SH}^2 < 4 \lambda_S \lambda_H$; 
\item The backreaction on the inflationary dynamics is negligible (Eq.~\eqref{eq:noBR}): $|\rho_H + \rho_{S}| < \rho_{\phi}$;
\item The $S$ field decays sufficiently rapidly after inflation (Eq.~\eqref{eq:sdecay}): $\lambda_S > 3172 \pi^3 c_S \left(\dfrac{H_\mathrm{inf}}{m_S}\right)^3$; 
\item The EFT remains valid (Eq.~\eqref{eq:EFT}):  $S_0, H_0 < \Lambda_W$.
\end{enumerate}

Note that condition~6, together with condition~3, typically implies  
a mass hierarchy between charged leptons and neutrinos, $m_{e_i} > m_{\nu_i}$, for $\lambda_S<\lambda_H$. 
Therefore, 
once condition~2
is satisfied for the neutrino masses, the corresponding condition for the charged leptons 
is automatically fulfilled.

\begin{figure}[htbp]
\centering
\includegraphics[width=0.7\textwidth]{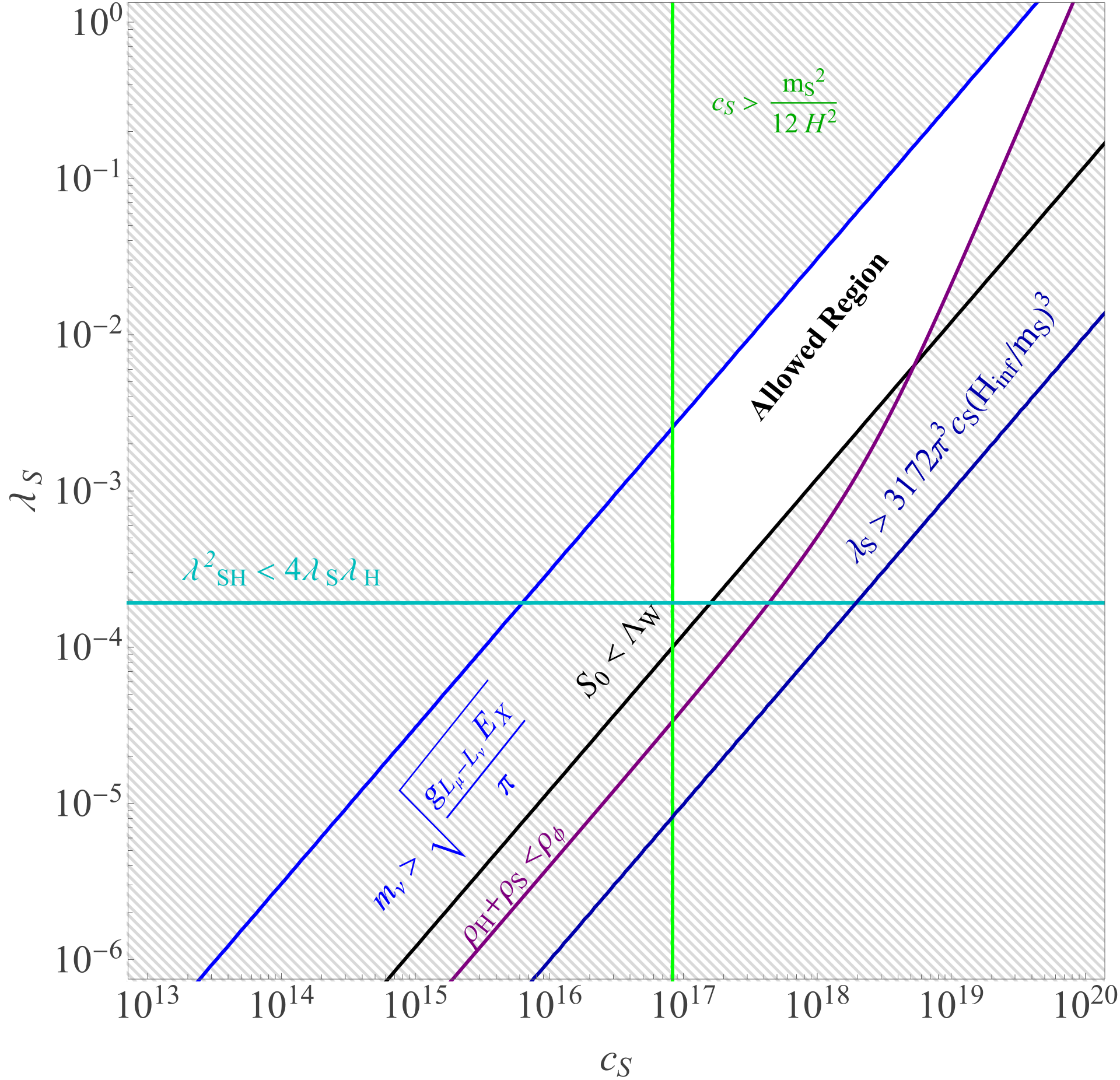}
\caption{\justifying Parameter space constraints for the spectator field model.  
The green line denotes the curvature-induced mass domination condition,  $c_S > \dfrac{m_S^2}{12H^2}$ with the region to the right allowed.
The blue line corresponds to the lower bound from the Schwinger suppression, $m_\nu > \sqrt{g_{L_i-L_j}E_X/\pi}$, where the region to the right is allowed. 
The cyan line represents the stabilization condition for a non-zero expectation value of the $S$ field,  $\lambda_{SH}^2 < 4 \lambda_S \lambda_H$, with the allowed region lying above the line.. 
The purple line indicates the backreaction constraint, $\rho_S+\rho_H < \rho_{\phi}$, allowing the region to the left.
The black line shows the EFT validity condition, $S_0< \Lambda_W$, with the allowed region to the left. 
The dark blue line represents the requirement that the $S$ field decays rapidly after inflation $\lambda_S > 3172 \pi^3 c_S \left(\dfrac{H_\mathrm{inf}}{m_S}\right)^3$, again allowing the region to the left.  
The white region denotes the parameter space simultaneously satisfying all constraints.}
\label{newmodel}
\end{figure}

Figure~\ref{newmodel} shows the region of parameter space in the $c_S$-$\lambda_S$ plane that satisfies all six conditions listed above. 
Throughout this figure, we fix $H_\mathrm{inf} =10^{-2}$ GeV, $\xi =17$, $g_{L_i-L_j} = 3 \times 10^{-4}$;  
the motivation for these choices will be discussed in the following sections. 
We also adopt the benchmark values 
$\lambda_{SH}=0.01, m_S= 10^7$ GeV, and $\Lambda_W= 10^9$ GeV. For this choice of parameters, we find $H_0<S_0$, and therefore once condition~6 is satisfied for $S_0$, the corresponding condition for $H_0$ is automatically fulfilled.
The electric field strength is taken to be $E_X = 1.6 \times 10^{-2} (\mathrm{e}^{\pi \xi}/\xi^{3/2}) H_\mathrm{inf}^2$ as given in Eq.~\eqref{eq:axion_EX}. 
Although a relatively large value of $c_S$ is required, 
this figure demonstrates  that the Schwinger effect can be effectively suppressed in this representative example. 
With this working example in hand, we next examine the post-inflationary cosmic history, assuming initial conditions in which 
helical $L_i-L_j$ magnetic fields are present, while no lepton asymmetry has yet been generated in the fermion sector.

\subsection{Survival of the $L_i-L_j$ magnetic field during the magnetohydrodynamic regime}

Next, we examine how the asymmetry remains stored in the magnetic helicity until the U(1)$_{L_i-L_j}$ symmetry breaking,  
working within the instantaneous reheating approximation. 
This is achieved if the coherence length of the $L_i-L_j$ magnetic fields is sufficiently large, so that  after inflation (reheating) their evolution can be described within 
with magnetohydrodynamics (MHD) and the non-linear effects prevents the magnetic fields (and helicity) from decaying. 
In the following, we present the necessary conditions for these requirements to be satisfied, 
based on the discussion in Ref.~\cite{Fukuda:2024pkh}. 

The situation is essentially the same as that for the SM hypercharge gauge field\,\cite{Giovannini:1997eg,Son:1998my,Jedamzik:1996wp,Banerjee:2004df,Domcke:2019mnd,Fukuda:2024pkh}. 
Let us therefore consider
the following set of MHD equations, 
\begin{align}
    \vec{J}_{L_i-L_j} &= \sigma (\vec{E}_X + \vec{v}\times\vec{B}_X), \label{eq:ohm}\\
    \nabla\times\vec{B}_X &= \vec{J}_{L_i-L_j} \label{eq:Ampere},\\
    \frac{\partial}{\partial t} \vec{v} + (\vec{v}\cdot\nabla)\vec{v} &= \frac{\eta_\mathrm{vis}}{\rho + p} \nabla^2 \vec{v} + \frac{1}{\rho + p}\qty(\vec{J}_{L_i-L_j}\times\vec{B}_X). \label{eq:navierstokes}
\end{align}
Here $\vec{J}_{L_i-L_j}$ is the total $L_i-L_j$ current, $\sigma$ represents the $L_i-L_j$ conductivity, $\vec{v}$ denotes the velocity of the fluid, $\eta_\mathrm{vis}$ is the shear viscosity, $\rho$ is the energy density of the fluid, and $p$ is the pressure of the fluid, respectively.  
Our analysis relies on several standard assumptions: the $L_i-L_j$ magnetic field is subdominant compared to the  radiation energy density of the plasma, $B_X^2 \ll \rho$; all particles can be treated as massless and are in thermal equilibrium; and the characteristic dynamical scale, $L$, is much larger than the mean free path of the $L_i-L_j$ charge carriers. 

The effect of cosmic expansion does not explicitly enter the MHD equations, since for a gauge theory without massive particles, we can eliminate it through a conformal transformation~\cite{Brandenburg:1996fc}. Following 
Ref.~\cite{Domcke:2019mnd}, we further assume that the magnetic and velocity fields are characterized by the same typical spatial scale. While this assumption is questioned by recent studies~\cite{Uchida:2022vue,Uchida:2024ude}, it is sufficient for our purposes to obtain a rough but reliable estimate.

Let us discuss the validity of the MHD equations, Eqs.~\eqref{eq:ohm}, \eqref{eq:Ampere}, and \eqref{eq:navierstokes}. 
Equation~\eqref{eq:ohm} represents the Ohm's law for the $L_i-L_j$ current. 
Its form can be qualitatively understood within the Drude model~\cite{Arnold:2000dr}, in which
the conductivity 
$\sigma$ in the conformal frame is estimated as
\begin{align}
    \sigma \simeq a(T) \sum_I q_{I}^2 g_{L_i-L_j}^2 g_I T^2 \tau_I. 
\end{align}
Here $T$ is the temperature of the universe, $a(T)$ is the corresponding scale factor at $T$, $q_I$ is the $L_i-L_j$ charge of particle $I$, $g_I$ is its number of internal degrees of freedom, and $\tau_I$ is its mean free time. 
For SM particles contents, 
the right-handed charged leptons
are the most weakly interacting particles charged under U(1)$_{L_i-L_j}$.
As a result, they possess
the longest mean-free time, $\tau \sim 1/(g'^4 T)$, where $g'$ is the $\text{U}(1)_Y$ gauge coupling constant. 
Consequently, their contribution dominates  the conductivity.
For the subsequent analysis, we therefore adopt the representative estimate, 
\begin{equation}
    \sigma \sim a(T) \frac{g_{L_i-L_j}^2 T}{g'^4}.
\end{equation}

Equation~\eqref{eq:Ampere} is Amp\`{e}re's law for the $L_i-L_j$ current, derived from the Maxwell equation, 
\begin{align}
    \nabla\times\vec{B}_X = \vec{J}_{L_i-L_j} + \frac{\partial}{\partial t} \vec{E}_X.
\end{align}
The displacement current, $\partial \vec{E}_X / \partial t$, can be neglected provided that  the following two conditions are satisfied, 
\begin{align}
    \frac{1}{\sigma T'} & \ll 1, \label{eq:assump_ampere1}\\
    |\vec{v}| \cdot \frac{L}{T'} & \ll 1 \label{eq:assump_ampere2},
\end{align}
as can be seen from Eq.~\eqref{eq:ohm}. 
Here $T'$ is the typical timescale of the dynamics.
For a magnetic field of inflationary origin, $T'$ corresponds to the conformal time and is given by $T' \simeq a(T) H(T_\text{re})^{-1} / a^2(T_\text{re})$. 
As a result, the first condition, Eq.~\eqref{eq:assump_ampere1}, can be rewritten as
\begin{align}
    \frac{g'^4}{g_{L_i-L_j}^2} \ll \frac{M_\mathrm{pl}}{T_\text{re}}. \label{eq:require1}
\end{align}
The second condition, Eq.~\eqref{eq:assump_ampere2}, is satisfied
for non-relativistic fluid velocities, $|\vec{v}| \ll 1$, a regime that will be justified in our subsequent analysis.

Equation~\eqref{eq:navierstokes}  is the Navier-Stokes equation governing the fluid dynamics of the system. 
Assuming a uniform fluid with constant energy density $\rho$ and pressure $p$, the shear viscosity, $\eta_{\mathrm{vis}}$, is given by~\cite{Weinberg:1971mx,Arnold:2000dr}
\begin{align}
    \eta_{\mathrm{vis}} = \frac{4}{15}\rho\tau, 
\end{align}
where $\tau$ denotes the mean free time of the plasma. 
The corresponding
kinetic viscosity is then defined as $\nu \equiv \eta_{\mathrm{vis}}/(\rho + p)$. 
For the SM plasma, this yields the estimate
\begin{align}
    \nu \sim g'^{-4} (a(T) T)^{-1}.
\end{align}

Now let us discuss the dynamics of the magnetic field and the fluid. 
It is well known that if the magnetic Reynolds number $R_m$ is sufficiently large,
\begin{align}
    R_m \equiv \sigma L |\vec{v}| \gg 1 \label{eq:assump_magrey},
\end{align}
the nonlinear term in the magnetic field evolution equation cannot be neglected. 
In this regime, Ohmic diffusion is suppressed and the magnetic helicity is well conserved~\cite{Giovannini:1997eg,Son:1998my,Jedamzik:1996wp,Banerjee:2004df,Domcke:2019mnd,Fukuda:2024pkh}. 
To obtain an estimate of $R_m$, we also need the typical fluid velocity. Although a precise determination requires numerical simulations, a rough estimate can be obtained using the Navier-Stokes equation. 
Following Refs.~\cite{Domcke:2019mnd,Fukuda:2024pkh}, we estimate the fluid velocity as
\begin{equation}
    |\vec{v}| \sim \left\{\begin{array}{ll} \dfrac{B_X}{T^2} & R_e \gg 1 \ \text{(turbulent regime)}, \\\sqrt{R_e}\dfrac{B_X}{T^2} \sim \dfrac{L}{\nu}\dfrac{B_X^2}{T^4} &  R_e \lesssim 1 \ \text{(viscous regime)}, 
    \end{array} \right.  
\end{equation}
where
\begin{align}
    R_e \equiv \frac{|\vec{v}| L}{\nu}, 
\end{align}
is the kinetic Reynolds number. 
In either regime, the velocity of the fluid is much smaller than unity, since the magnetic field energy density is subdominant compared to the radiation energy density of the plasma,  $B_X^2 \ll \rho$. This ensures that the condition in Eq.\,\eqref{eq:assump_ampere2} is satisfied.

Let us summarize the conditions required for the survival of 
the magnetic helicity. 
There are three such conditions: 
a large electrical conductivity (Eq.~\eqref{eq:assump_ampere1}), 
a small fluid velocity (Eq.~\eqref{eq:assump_ampere2}), 
and a large magnetic Reynolds number (Eq.~\eqref{eq:assump_magrey}). 
The first two conditions are readily satisfied. The high conductivity requirement is met provided 
the $L_i-L_j$ gauge coupling is not extremely small, 
depending on
the reheating temperature (Eq.~\eqref{eq:require1}),  
while the small-velocity condition is automatically fulfilled since the magnetic field energy density is subdominant compared to the radiation energy density  of the plasma. 
Assuming the magnetic field was generated during inflation, the third condition, namely a sufficiently large magnetic Reynolds number, can be expressed as
\begin{align}
    \frac{g_{L_i-L_j}^2B_{X0} H(T_\text{re})^{-1}}{g'^4T_\text{re}} \times \min\qty(1, \frac{g'^4 B_{X0} H(T_\text{re})^{-1}}{T_\text{re}}) \gg 1, \label{infmagturb}
\end{align}
where $B_{X0}$ is the amplitude of the magnetic field at the end of inflation and we have taken $L(T_\mathrm{re}) \simeq (a(T_\mathrm{re}) H_\mathrm{re})^{-1}$.

Figure~\ref{withoutschwinger} displays the parameter space satisfying Eq.~\eqref{infmagturb} in the $\eta_\mathrm{H}-H_\mathrm{inf}$ plane, 
together with the backreaction constraint that the magnetic field energy density remain subdominant to that of inflaton, $\rho_X < 0.01 \rho_\phi$ (Eq.~\eqref{eq:negbr}). 
Here we fix $Q=3, g_{L_i-L_j} = 3\times 10^{-4}$. 
In this figure, we do not take into account the contribution of fermions produced via the Schwinger effect, assuming it is effectively switched off. 
For  smaller values of $\eta_\mathrm{H}$, the magnetic helicity is dissipated, while for larger $\eta_\mathrm{H}$ the gauge field energy density becomes excessively large and violates the backreaction bound.
Nevertheless, a viable parameter region exists in which magnetic fields are consistently generated during inflation and survive without  dissipation, in particular for relatively small values of the Hubble parameter during inflation. 
Remarkably, within this allowed region, a large magnetic helicity can be produced to account for the reduced helium abundance suggested by the ACT and SPT. 
For reference, the contours of the instability parameter $\xi$ are shown; 
the  parameter regions favored by the ACT and SPT correspond to $\xi>16$. 
In the next subsection, we discuss the condition under which the lepton asymmetry is appropriately generated after the EWPT,  which in turn justifies the choice of the gauge coupling, $g_{L_i-L_j} = 3\times 10^{-4}$.

\begin{figure}[htbp]
\centering
\includegraphics[width=0.7\textwidth]{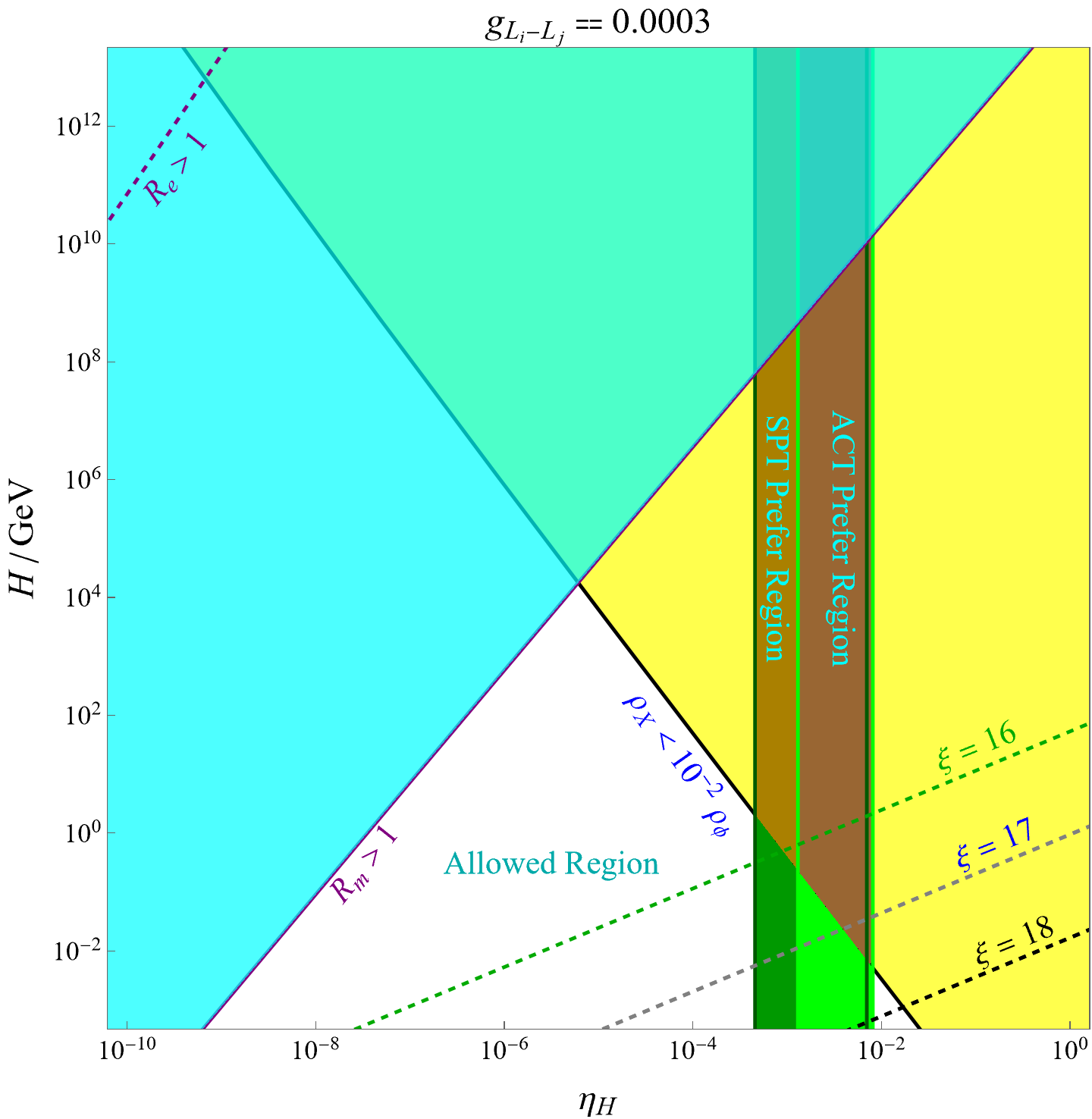}
\caption{
    \justifying
    The parameter space constrained by the survival condition of magnetic helicity after inflation, in the absence of the Schwinger effect, is shown in the $\eta_\mathrm{H}$-$H_\mathrm{inf}$ plane for $g_{L_i-L_j} =3\times 10^{-4}$. 
    The purple solid line correspond to $R_m=1$, 
    and the cyan-shaded regions to the left is excluded by the condition $R_m<1$.
    The black solid line indicates $\rho_{X}=0.01\rho_{\phi}$,  
    while the yellow-shaded regions to the right is excluded by non-negligible backreaction from the energy density of the gauge fields during inflation. The brown and green shaded regions represent the parameter space preferred by the ACT and SPT results, assuming the corresponding electron neutrino asymmetry is generated~(Eq.~\eqref{eq:nueasymACT}). The green-shaded regions denote the parameter space accessible in our scenario. The green, gray, and black dashed lines show contours of $\xi=16,17,18$, respectively.  }
\label{withoutschwinger}
\end{figure}

\subsection{Magnetic helicity decay and leptogenesis}

Finally, we examine how the U(1)$_{L_i-L_j}$ symmetry is broken and how the magnetic 
helicity decay leads to the  generation of a lepton asymmetry. 
Here we suppose that it is spontaneously broken by the Higgs mechanism, so that the U(1)$_{L_i-L_j}$ gauge field acquires a mass given by 
\begin{equation}
    m_{Z'} = q g_{L_i-L_j} v_{L_i-L_j} (T),  
\end{equation}
where $q$ is the charge of the U(1)$_{L_i-L_j}$-breaking Higgs field and $v_{L_i-L_j} (T)$ is its temperature dependent expectation value. 
Assuming that the Higgs potential takes the form,
\begin{equation}
    V(h;T)=\left(\frac{q^2 g_{L_i-L_j}^2}{4} T^2 - |m_{h_X}^2|\right) |h|^2+ \lambda_{h_X} |h|^4, 
\end{equation}
with $\lambda_{h_X}\sim (q g_{B-L})^2$, 
the U(1)$_{L_i-L_j}$ symmetry is broken at at the critical temperature
\begin{equation}
    T_c \sim \frac{|m_{h_X}|}{q g_{L_i-L_j}} \sim v_{L_i-L_j}(T=0) \sim \frac{m_{Z'}}{q g_{L_i-L_j}}. 
\end{equation}

As argued in Ref.~\cite{Fukuda:2024pkh}, in the Higgs phase the magnetic field and its helicity decay either through the imaginary part of the gauge-field self-energy or via Ohmic dissipation in the Higgs phase. 
Both decay time scales, $\tau_X \sim (24 \pi/ g_{L_i-L_j}^2)/m_{Z'}$ and $\tau_\mathrm{MHD} \sim \sigma/m_{Z'}^2$, are much shorter than the Hubble time at the epoch of symmetry breaking. 
Therefore, the magnetic helicity is expected to decay promptly after the U(1)$_{L_i-L_j}$ symmetry breaking. 
Since the anomaly equation remains unchanged in the presence of a non-zero expectation values of the Higgs field, 
$L_i$ and $L_j$ asymmetries are generated through the transfer of the magnetic helicity into the fermion sector, as described by Eqs.~\eqref{eq:csconservation}. 
Quantitatively, the resulting flavored lepton-to-entropy ratios are given by
\begin{equation}
     \frac{n_{L_i}}{s} =\frac{n_{L_j}}{s} = \eta_\mathrm{H}. \label{eq:lepflasym}
\end{equation}

Yet another possibility, which was not considered in Ref.~\cite{Fukuda:2024pkh}, is that the helical magnetic fields are stored in the cosmic strings after symmetry breaking. 
In this case, however, the helicity would be carried by  linked cosmic strings, whose typical length scale is much smaller than the Hubble scale. 
After the symmetry breaking and and the cosmic string formation, 
reconnection processes take place efficiently as the system quickly enters the scaling regime of the cosmic string network. 
Through such reconnections, magnetic helicity is expected to disappear promptly. 
Unless the cosmic strings are destroyed via monopole anti-monopole pair production, the anomaly equation remains valid during the magnetic helicity decay through the reconnection~\cite{Hamada:2025cwu,Fukuda:2025nmc}, 
which is the situation relevant to the present study. 
Therefore, our conclusion remains unchanged: {\it immediately after the U(1)$_{L_i-L_j}$ symmetry breaking $L_i$ and $L_j$ asymmetries are generated in the Universe with the magnitude estimated in Eq.~\eqref{eq:lepflasym}}.  

For the scenario considered here to be viable, 
the generation of $L_i$ and $L_j$ asymmetries -- and hence the U(1)$_{L_i-L_j}$ symmetry breaking -- must occur after the EWPT and before the BBN, or before the onset of the neutrino oscillations in the case of $L_\mu-L_\tau$. 
This requirement translates into a constraint on  the gauge coupling $g_{L_i-L_j}$, 
\begin{equation}
    T_\mathrm{BBN} < T_c <T_\mathrm{EW} \Rightarrow \frac{q m_{Z'}}{T_\mathrm{EW}} < g_{L_i-L_j} < \frac{q m_{Z'}}{T_\mathrm{BBN}}, \label{eq:gcbbnew}
\end{equation}
where $T_\mathrm{EW} \simeq 160$ GeV and $T_\mathrm{BBN} \simeq 10$ MeV. 
Assuming  $q=\mathcal{O}(1)$, 
current observational and experimental constraints on 
U(1)$_{L_e-L_\mu}$ and U(1)$_{L_e-L_\tau}$ gauge bosons do not 
allow sufficiently large gauge coupling to satisfy the condition in Eq.~\eqref{eq:higgsexp}~\cite{Bauer:2018onh}.  
In contrast, for the U(1)$_{L_\mu-L_\tau}$ gauge boson, the constraint around $m_{Z'} \sim 10$ MeV is significantly weaker.  
In this mass range,  the most stringent bound is provided by the fixed-target experiment NA64$\mu$, which gives $g_{L_\mu-L_\tau}<6\times 10^{-4}$~\cite{NA64:2024klw}. This upper bound is compatible with  the lower bound implied by Eq.~\eqref{eq:gcbbnew}.
This consideration motivates our benchmark choice $g_{L_\mu-L_\tau} = 3\times 10^{-4}$. 
Interestingly, a gauge coupling of this magnitude  is also favored in  gauged U(1)$_{L_\mu-L_\tau}$ models that aim to explain~\cite{He:1990pn,Baek:2001kca,Lindner:2016bgg} the possible muon $g-2$ anomaly~\cite{Muong-2:2021ojo,Muong-2:2024hpx,Aoyama:2020ynm} (see, however, Ref.~\cite{Aliberti:2025beg}). 
Figure~\ref{fig: g_L_mu-L_tau VS m_X} shows the parameter region  in the $m_{Z'}$-$g_{L_\mu-L_\tau}$ plane where the U(1)$_{L_\mu-L_\tau}$ is broken after the EWPT and before the BBN, 
together with 
\begin{figure}[htbp]
\centering
\includegraphics[width=0.7\textwidth]{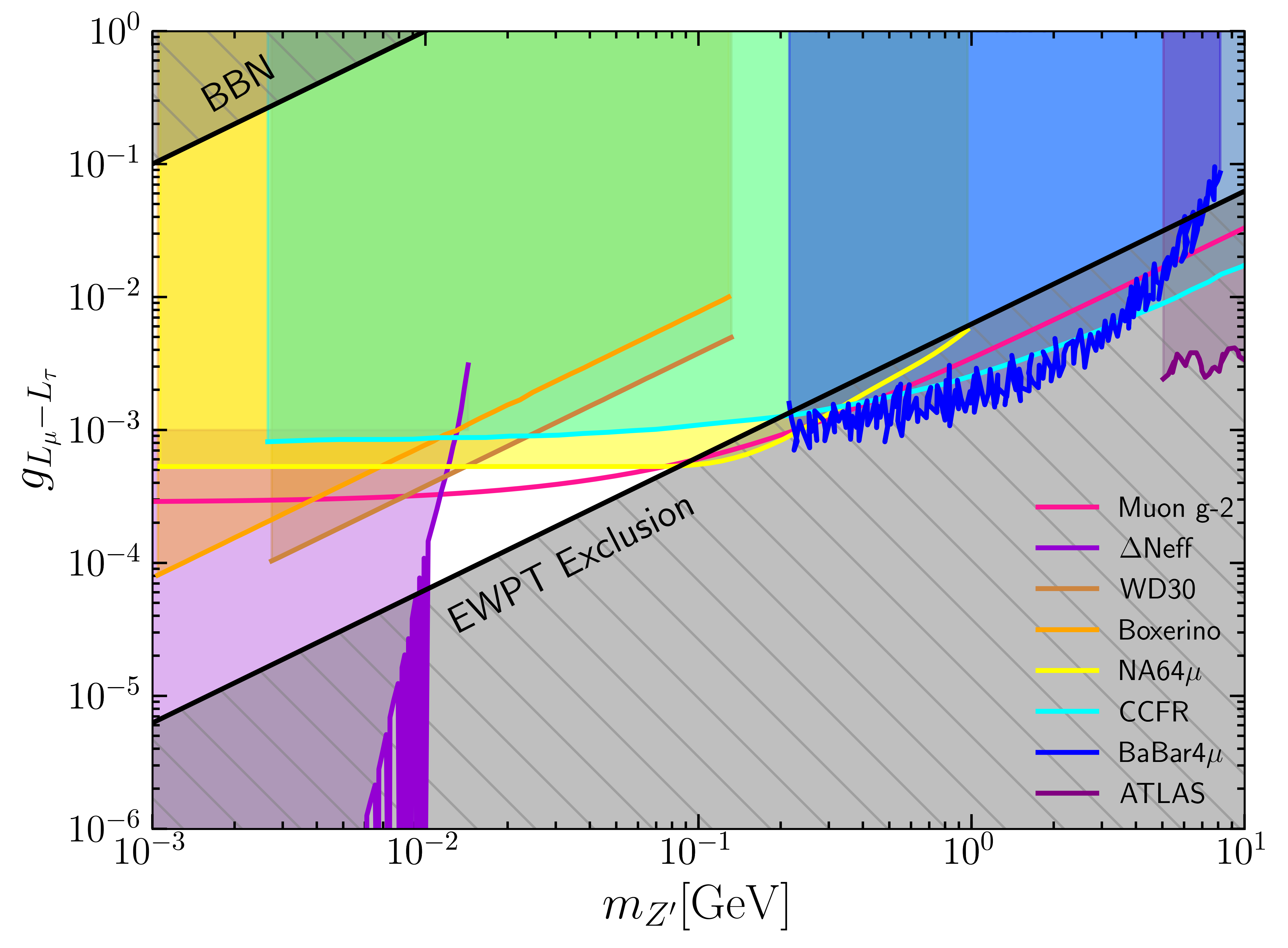}
\caption{
    \justifying 
    The parameter space in the $m_{Z'}$-$g_{L_\mu-L_\tau}$  plane in which the  the U(1)$_{L_\mu-L_\tau}$ symmetry breaking to occur after EWPT and before BBN is shown,  together with existing experimental constraints. The upper gray-shadowed region is excluded by the condition $T_{\mathrm{BBN}}<T_c$, while the lower gray-shadowed region is excluded by $T_c<T_{\mathrm{EW}}$, so that remaining region satisfies Eq.~\eqref{eq:gcbbnew}. 
    Colored regions indicate experimental constrains for $Z'$ parameter space: the bound from the effective number of neutrino species $\Delta N_\mathrm{eff}\leq0.5$, for $g_{\mu\tau}\geq4\times10^{-9}$ and $m_{Z'}\leq0.1\mathrm{GeV}$ (light purple) \cite{Escudero:2019gzq}; ATLAS searches (thick purple) \cite{ATLAS:2023vxg,ATLAS:2024uvu};  hot white dwarf cooling (thick orange) \cite{Bell:2025acg,Foldenauer:2024cdp}; the muon anomalous magnetic moment $\Delta a_\mu$ 
    within $1\sigma$ uncertainty (pink) \cite{Muong-2:2025xyk,Aliberti:2025beg}; Borexino (light orange) \cite{Bellini:2011rx}; NA64$\mu$ (yellow) \cite{Andreev:2024lps}; BaBar \& Bell (thick blue) \cite{BaBar:2016sci,Belle-II:2024wtd}; and  neutrino trident production from CCFR (light blue) \cite{Altmannshofer:2014pba}. For the experimental constraints, we use the compiled data in \cite{Bernal:2025szh}.}
\label{fig: g_L_mu-L_tau VS m_X}
\end{figure}
existing 
experimental constraints. 
Although the  experimental constraints are rather restrictive, we find that non-vanishing  parameter region remains viable. 
Note that heavier mass range of gauge boson mass is ruled out by, such as,  ATLAS~\cite{ATLAS:2024uvu} and CCFR~\cite{Altmannshofer:2014pba}.

\ 

In summary, we have examined a scenario in which axion inflation coupled to a U(1)$_{L_i-L_j}$ gauge boson results in a sizable electron neutrino asymmetry at  BBN, leading to a reduced primordial helium abundance favored by recent results from ACT, SPT and EMPRESS. 
In the case of a gauged U(1)$_{L_\mu-L_\tau}$ theory, we have successfully constructed a model that generates sufficiently large helical ${L_\mu-L_\tau}$ magnetic field while effectively switching off the Schwinger effect. 
For the benchmark parameters, $H_\mathrm{inf} \simeq 10^{-2}$ GeV, $\xi \simeq 17$, $g_{L_\mu-L_\tau} \simeq 3\times 10^{-4}$, and $m_{Z'} \simeq 10$ 
MeV, the helical magnetic fields enter the hydrodynamic regime, where the magnetic helicity is stored in the magnetic sector.  
The helicity subsequently decays after EWPT and before the BBN thereby  
generating $L_\mu$ and $L_\tau$ asymmetries. 
The electron neutrino asymmetry is then induced through the neutrino oscillations, as described in Eq.~\eqref{eq:neoscillation}, 
and can account for the helium abundance suggested by  ACT and SPT.  
One might be concerned that, during the MHD regime, lepton asymmetry could be partially generated through the magnetic dissipation, potentially leading to an overproduction of  baryon asymmetry via sphaleron process. 
We have confirmed, however, that in the parameter of  interest,  the relatively relatively large coherence length of the magnetic field sufficiently suppress such dissipation, preventing excessive baryon asymmetry production.

\section{Conclusion and discussion \label{sec:summary}}

In this paper, we have investigated lepton number generation from axion inflation. 
We considered a gauged lepton flavor symmetry, U(1)$_{L_i-L_j}$,  whose gauge field has a Chern--Simons coupling to the inflaton, leading to a tachyonic instability of one circular polarization mode during inflation.
Through the chiral anomaly, the generation of such helical gauge field is associated with the lepton number production. 
Even setting aside the baryon overproduction problem, we found an upper bound of the generated lepton asymmetry arising from the backreaction of the produced gauge fields and fermions on the inflationary dynamics. 
By implementing a mechanism to suppress fermion production from the Schwinger effect during inflation, a much larger magnetic helicity can be generated. 
As a consequence, the lepton asymmetry is generated in the Universe at the time when the U(1)$_{L_i-L_j}$ gauge symmetry is broken. 

We find that, in the case of a gauged U(1)$_{L_\mu-L_\tau}$ symmetry, 
the symmetry breaking and the associated lepton number generation can occur later than the EWPT, thereby avoiding the baryon overproduction via the sphaleron conversion. 
The resulting lepton asymmetry can be sufficiently large to induce an electron neutrino asymmetry that reduces the helium abundance, in a manner compatible with recent CMB observations such as those by ACT and SPT, for inflationary parameters, $H_\mathrm{inf} \sim 0.01$ GeV and $\xi\sim 17$, at the end of inflation. 
Remarkably, the parameter range of the U(1)$_{L_\mu-L_\tau}$ symmetry relevant for this scenario is also compatible with interpretations of the possible muon $g-2$ anomaly. 

One may wonder whether such a low Hubble parameter and large value of the parameter $\xi$ are observationally and theoretically viable. 
While large values of $\xi \gtrsim 1.92 \text{--} 2.45$ are constrained by CMB observations on the scales corresponding to modes that exited the horizon  $\mathcal{N}_e \sim 50 -60$ e-folds before the end of inflation~\cite{Barnaby:2011vw,Barnaby:2011qe,Meerburg:2012id,Pajer:2013fsa}, 
the value required in our scenario, $\xi \sim 17$, corresponds to the modes generated near the end of inflation. 
Therefore, these constraints are generically irrelevant for magnetogenesis, as long as the inflaton potential  is not specified explicitly. 
Similarly, such a low Hubble parameter is  not realized in the simplest inflationary models, such as the $R^2$ inflation or Higgs inflation, nor in the conventional axion inflation with a cosine-type potential. 
However, the shape of axion potential depends on the details of the underlying theory~\cite{Svrcek:2006yi,Silverstein:2008sg,McAllister:2008hb,Burgess:2014oma,Nomura:2017ehb}. 
Conversely, our scenario provides a motivation to explore axion inflation models that predict a lower Hubble parameters during inflation.
Note that, in the absence of a detection of the inflationary CMB $B$-mode polarization, the Hubble parameter during inflation is not directly constrained provided that the reheating temperature is compatible with BBN. 
Indeed, inflationary models with a low Hubble parameter have been proposed in literature; see, for example, Ref.~\cite{Takahashi:2018tdu}. 
A concrete realization of our scenario, including an explicit construction of the inflaton sector, is left for future study.

Observationally and experimentally, our scenario could be tested by search for the  U(1)$_{L_\mu-L_\tau}$ gauge boson, 
as well as by future measurements of gravitational waves sourced by gauge field produced during inflation. 
While the basic ideas of our scenario are robust, 
our discussion is based primarily on order-of-magnitude estimates, 
and a more quantitative treatment will be necessary to fully confront upcoming experimental data. 
One of the most important open issues is the detailed dynamics of the U(1)$_{L_\mu-L_\tau}$ symmetry breaking and the associated generation of  lepton number. 
In this work, we have considered three possible mechanisms for lepton asymmetry generation around the symmetry breaking: decay through the imaginary part of the gauge-field self energy,  Ohmic dissipation; and reconnection processes of the cosmic strings. 
Although all of these mechanisms generally  predict rapid lepton number generation immediately after symmetry breaking, 
the precise timing and efficiency of the process are sensitive to the underlying dynamics. 
It would therefore be highly interesting to investigate this problem using numerical lattice simulations or MHD simulations of symmetry breaking in the presence of helical magnetic fields.

\acknowledgments
The authors are grateful to Hajime Fukuda for collaboration during the early stages of this study. 
They also thank Valerie Domcke, Eiichiro Komatsu, Kyohei Mukaida, Mikhail Shaposhnikov, and Anna Tokareva for useful comments and discussions. 
The work of K.K. was supported by the National Natural Science Foundation of China (NSFC) under Grant No.~12347103 and by JSPS KAKENHI Grant-in-Aid for Challenging Research (Exploratory) JP23K17687.

% bib
\bibliographystyle{utphys}
\bibliography{papers}

\end{document}